\documentclass[a4paper,11pt]{article}
\pdfoutput=1
\usepackage{jheppub}
\usepackage{amsmath}
\usepackage{braket}
\usepackage{hyperref}
\usepackage{slashed}
\usepackage{color,soul}
\usepackage{enumitem}
\allowdisplaybreaks
\def\bea#1\eea{\begin{align}#1\end{align}}
\newcommand{\nnu}{\nonumber\\}
\newcommand{\nn}{\nonumber}

\newcommand{\mcdot}{\!\cdot\!}
\newcommand{\bef}{\begin{figure}[hbt]\centering}
\newcommand{\eef}{\end{figure}}

\newcommand{\vc}[1]{\boldsymbol{#1}}
\newcommand{\f}{\frac}

\def \be  {\begin{equation}}
\def \ee  {\end{equation}}
\def \ba  {\begin{eqnarray}}
\def \ea  {\end{eqnarray}}
\newcommand\as{\alpha_s}
\newcommand{\vect}[1]{\boldsymbol{#1}} 

\title{Effective field theory approach to open heavy flavor production in heavy-ion collisions}

\author{Zhong-Bo Kang,}
\author{Felix Ringer}
\author{and Ivan Vitev}

\affiliation{Theoretical Division, 
                 Los Alamos National Laboratory,
                 Los Alamos, NM 87545, USA}
                                         
\emailAdd{zkang@lanl.gov}
\emailAdd{f.ringer@lanl.gov}
\emailAdd{ivitev@lanl.gov}

\abstract{We develop a version of Soft Collinear Effective Theory (SCET) which includes finite quark masses, as well as Glauber gluons that describe the interaction of collinear partons with QCD matter. In the framework of this new effective field theory,  labeled SCET$_{\mathrm{M,G}}$,  we derive the massive splitting functions in the vacuum and the QCD medium for the processes $Q\to Qg$, $Q\to gQ$ and $g\to Q\bar Q$. The numerical effects due to finite quark masses are sizable and our results are consistent with the traditional approach to parton energy loss in the soft gluon emission limit. In addition, we present a new framework for including the medium-induced full splitting functions consistent with next-to-leading order calculations in QCD for inclusive hadron production. Finally, we show numerical results for the suppression of  $D$- and $B$-mesons in heavy ion collisions at $\sqrt{s_{\mathrm{NN}}}=5.02$~TeV and 2.76~TeV and compare to available data from the LHC. }

\begin{document}
\maketitle

\section{Introduction}
\label{sec:one}

Inclusive open heavy flavor production in proton-proton and heavy-ion collisions is considered to be one of the most important tests of our understanding of QCD and, in particular, of quark mass effects both in the vacuum and in a QCD medium. Measurements of heavy flavor meson cross sections have been performed at the Tevatron~\cite{Acosta:2003ax,Acosta:2001rz,Acosta:2004yw} and at the Relativistic Heavy Ion Collider (RHIC)~\cite{Adare:2010de, Adamczyk:2014uip, Adare:2014rly}. More recently, the ATLAS, CMS and ALICE experimental collaborations at the Large Hadron Collider (LHC) have provided high precision data~\cite{Khachatryan:2011mk,Khachatryan:2016ypw,Grelli:2012yv,ATLAS:2013cia,Aad:2015zix,CMS:2015ixs,CMS:2016nrh,Adam:2016ich} for several center of mass (CM) energies, and they will continue to extend the currently existing suite of data sets with future measurements. For a recent review on heavy flavor production, see Ref.~\cite{Andronic:2015wma}.

Theoretical and experimental advances in understanding the nuclear modification of  light hadrons, heavy mesons, as well as jets and jet substructure in nucleus-nucleus reactions have been a highlight of the heavy ion programs at RHIC and the LHC~\cite{Akiba:2015jwa}. Such highly energetic particles and jets are powerful and valuable probes of the quark-gluon plasma (QGP) produced in these collisions. In particular, open heavy flavor production plays a crucial role in elucidating the properties of QGP and has received growing attention from the experimental and theoretical communities in recent years. Earlier data from RHIC and preliminary measurements from the CMS collaboration~\cite{CMS:2016nrh} at $\sqrt{s_\mathrm{NN}}=5.02$~TeV show that the suppression rates for $D^0$ mesons are in fact the same as for light charged hadrons within the experimental uncertainty, contrary to the early expectation from traditional parton energy loss framework~\cite{Dokshitzer:2001zm}. This has stimulated a series of work addressing the interaction of heavy quarks with the QCD medium in the literature, see e.g.~\cite{Djordjevic:2003zk,Armesto:2005mz,Wicks:2005gt,Adil:2006ra,Vitev:2006bi,Kang:2011rt,Huang:2013vaa,He:2014cla,Ozvenchuk:2014rpa,Das:2015ana,Xu:2015bbz,Huang:2015mva,Song:2015sfa,Djordjevic:2015hra,Cao:2015kvb,Cao:2016gvr} and references therein. Different than light hadrons, the heavy quark mass introduces an additional perturbative scale besides the large transverse momentum $p_T$ at which the heavy meson is produced. This feature makes the description of heavy quark dynamics both challenging and also particularly interesting, as it can reveal unique information about the QCD medium. 

In the past few year, there has been a new development in the theoretical description of observables in heavy-ion collisions. In~\cite{Idilbi:2008vm,Ovanesyan:2011xy}, the powerful techniques of Soft Collinear Effective Theory (SCET)~\cite{Bauer:2000yr,Bauer:2001yt,Beneke:2002ph} were first applied to describe the interactions of hard probes (highly energetic particles and jets) with the QCD medium. The underlying idea is to include a Glauber mode that describes the interaction of highly energetic partons with the QCD medium. The effective field theory describing the medium interactions is commonly referred to as SCET$_\mathrm{G}$. In~\cite{Ovanesyan:2011xy,Ovanesyan:2011kn,Fickinger:2013xwa,Ovanesyan:2015dop} the in-medium splitting functions were derived to first order in opacity. By taking the soft emission limit, the full in-medium splitting functions reduce to the results obtained within traditional approaches to parton energy loss~\cite{Gyulassy:2000fs,Gyulassy:2000er}. Using SCET$_\mathrm{G}$ allows to systematically go beyond these traditional approaches. In~\cite{Kang:2014xsa,Chien:2015vja,Chien:2015hda,jet_frag} several applications of the in-medium splitting functions were developed. In~\cite{Kang:2014xsa,Chien:2015vja} a medium-modified DGLAP evolution approach was  successfully applied to describe the suppression of light charged hadrons in the medium. In~\cite{Chien:2015hda,jet_frag} the SCET$_\mathrm{G}$ based splitting functions were used to describe both inclusive jet production and jet substructure observables in heavy ion collisions.

In this paper, we perform the next logical step in this line of work by including finite mass effects in the SCET$_\mathrm{G}$ Lagrangian and, thus, enable the effective theory study the of interactions of heavy quarks with the QCD medium. The SCET Lagrangian in the vacuum with quark masses was first derived in~\cite{Leibovich:2003jd}. The corresponding theory in the vacuum is commonly referred to as SCET$_\mathrm{M}$. Consequently, we label the new effective field theory presented in this work SCET$_{\mathrm{M,G}}$. With this new theory at hand, we extend the in-medium splitting functions to the massive case. The newly derived results can be used to describe the suppression of open heavy flavor production in heavy-ion collisions. We introduce a new way to implement the in-medium corrections consistently at next-to-leading order (NLO) in perturbative QCD. This can be achieved by formally introducing medium-modified fragmentation functions based on the SCET$_{\mathrm{M,G}}$ splitting functions. We present first numerical results in this work and compare to data taken at the LHC. As it turns out, the description of the underlying proton-proton baseline plays an important role. Several different approaches are available in the literature to deal with heavy quark masses in the fragmentation process~\cite{Nason:1989zy,Nason:1993xx,Cacciari:1998it,Cacciari:2001td,Kneesch:2007ey,Kniehl:2008zza,Bauer:2013bza,Fickinger:2016rfd}. The suppression rates in heavy-ion collisions crucially depend on whether the heavy meson is produced by a fragmenting heavy quark or a gluon. Gluons lose more energy than heavy quarks when they undergo multiple scatterings and splittings in the medium before they eventually fragment into the observed heavy meson. In this work we analyze the different production mechanisms and study the associated suppression rates in Pb+Pb collisions.

The remainder of this paper is organized as follows. In section~\ref{sec:two}, we derive the Lagrangian of the new effective field theory SCET$_{\mathrm{M,G}}$ that includes both finite quark masses and Glauber gluons that describe the interaction with the QCD medium. We derive the massive vacuum and in-medium splitting functions and make the connection to previous results in the literature in the soft emission limit. In addition, we study the numerical impact of finite quark masses. In section~\ref{sec:three}, we start by introducing our new formalism that treats both vacuum and in-medium corrections consistently to NLO in QCD. We present numerical results for both proton-proton and heavy-ion collisions and compare to currently available data from the LHC. In section~\ref{sec:four}, we conclude and give an outlook.

\section{Effective field theory for massive quarks in the medium \label{sec:two}}

In this section, we first introduce the basic SCET ingredients relevant to our calculation. We then derive the Lagrangian for the effective theory SCET$_{\mathrm{M,G}}$, which describes the interactions of collinear heavy quarks with the QCD medium through Glauber gluon exchange. We use this new version of SCET to derive the massive splitting kernels in the vacuum and  in nuclear matter for the splitting processes $Q\to Qg$, $Q\to gQ$ and $g\to Q\bar Q$, where $Q$ represents a heavy quark. In order to establish the connection between our newly derived in-medium splitting functions and the results derived previously within the traditional parton energy loss approach, we further consider the limit of the splitting kernels where the emitted parton becomes soft. Finally, we present numerical results for the splitting functions and compare to the massless case~\cite{Ovanesyan:2011kn}, as well as the results from traditional parton energy loss calculations.

\subsection{Basic SCET ingredients}

SCET~\cite{Bauer:2000yr,Bauer:2001yt,Beneke:2002ph} is an effective field theory describing the dynamics of soft and collinear quarks and gluons in the presence of hard interactions. SCET has been applied successfully to hard-scattering processes at the LHC, in particular to the production of highly energetic hadrons and jets. We adopt the following convention for the light-cone notation.  We define two reference vectors $n^\mu=(1,0,0,1)$, $\bar n^\mu=(1,0,0,-1)$ satisfying $n^2 = \bar n^2 = 0$ and $n\cdot \bar n = 2$. Any four-vector $p^\mu$ can be written as $p^\mu=(p^+,p^-,\vect{p}_\perp)=(\bar n\cdot p,n\cdot p,\vect{p}_\perp)$. In other words,
\bea
p^\mu = p^+ \frac{n^\mu}{2} + p^- \frac{\bar n^\mu}{2} + p_\perp^\mu. 
\eea
The hierarchy between the hard, collinear and the soft scale is determined by the SCET power counting parameter $\lambda$. The SCET degrees of freedom have the following momentum scaling using light-cone coordinates $p^\mu= (p^+, p^-, \vect{p}_\perp) \sim p^+ (1, \lambda^2, \lambda)$ for a collinear mode and $p^\mu \sim p^+ (\lambda^2, \lambda^2, \lambda^2)$ for a soft mode. Note that in our notation, $p^+$ is a hard scale. The Glauber modes that we consider in the next section scale as $p^\mu \sim p^+ (\lambda^2, \lambda^2, \lambda)$. For a collinear quark, one separates the momentum as $p=\tilde p+k$, where $\tilde p=\bar n( n\cdot p)/2+p_\perp$ is the large label momentum and $k$ is the residual momentum. The label momentum component is removed by defining a quark field $\psi_{n,p}(x)$ through
\be\label{eq:psi1}
\psi(x)=\sum_{\tilde p}e^{-i\tilde p\cdot x}\psi_{n,p}(x) \, ,
\ee
where $\psi(x)$ is the standard QCD quark field. The four component field $\psi_{n,p}(x)$ has two large components $\xi_{n,p}(x)$ and two small components $\xi_{\bar n,p}(x)$. One defines the following two projections
\be\label{eq:psi2}
\xi_{n,p}(x)=\f{\slashed{n}\slashed{\bar n}}{4}\psi_{n,p}(x) \, , \quad \xi_{\bar n,p}(x)=\f{\slashed{\bar n}\slashed{ n}}{4}\psi_{n,p}(x) \, ,
\ee
with $\psi_{n,p}(x)=\xi_{n,p}(x)+\xi_{\bar n,p}(x)$. Operators in SCET are defined in terms of gauge invariant quark and gluon fields which are given by
\bea\label{eq:chiB}
\chi_{n} =  W_n^\dagger \xi_n,
\qquad
{\mathcal B}_{n\perp}^\mu =  \frac{1}{g}\left[W_n^\dagger iD_{n\perp}^\mu W_n\right],
\eea
with $iD_{n\perp}^\mu = {\mathcal P}_{n\perp}^\mu + gA_{n\perp}^\mu$ and ${\mathcal P}^\mu$ is the label momentum operator. Furthermore, $W_n$ is a Wilson line of collinear gluons,
\bea
W_n(x) = \sum_{\rm perms} \exp\left[-g\frac{1}{\bar{\cal P}} \bar n\cdot A_n(x)\right] \, .
\eea
with $\bar{\cal P} = \bar n\cdot {\mathcal P}$.

\subsection{SCET with quark masses and Glauber gluons\label{sec:SCETMG}}

When an energetic parton traverses a  dense and/or hot QCD medium, as produced in heavy-ion collisions, the formation of an in-medium parton shower can be described using perturbative methods. Following~\cite{Gyulassy:1993hr}, the medium can be thought of as color-screened quasi-particles that generate a Coulomb-like potential that effectively leads to a background field for the partons traveling through the QCD medium. The energetic parton that eventually produces a jet of particles undergoes multiple elastic interactions with the quasi-particles. In the vacuum, the parton shower forms by standard soft and collinear splittings. These processes are described by the original SCET Lagrangian. In the medium, one needs to consider additional medium-induced splitting processes. The interaction of collinear {\it massless} quarks and gluons with $t$-channel off-shell gluons is described by the SCET$_\mathrm{G}$ Lagrangian as derived in~\cite{Idilbi:2008vm,Ovanesyan:2011xy}. So far, SCET$_\mathrm{G}$ only contains the interaction of collinear quarks and gluons with the medium through the so-called Glauber gluon exchange, which has the momentum scaling $\sim p^+(\lambda^2,\lambda^2,\lambda)$. The corresponding soft sector has not been derived yet, nor the interactions between the Glauber gluons and soft gluons and quarks~\cite{Rothstein:2016bsq}. In general, Glauber modes play an essential role in various aspects of QCD, see for example~\cite{Bauer:2010cc,Fleming:2014rea,Rothstein:2016bsq} for more details. We leave the complete formulation of SCET$_\mathrm{G}$ with soft modes in the medium for future work. 

The purpose of this section is to derive an extension of the effective field theory SCET$_\mathrm{G}$, as presented in \cite{Ovanesyan:2011xy} for massless quarks and gluons, by including the effects of heavy quark masses. The resulting new version of the effective field theory, labeled  SCET$_\mathrm{M,G}$, will enable us to study the interactions of energetic heavy quarks with a hot QCD medium as a first example. We start by reviewing the previous work~\cite{Ovanesyan:2011xy}, in which several source fields and gauges were considered and the scaling of the in-medium background field was derived. The structure of the SCET$_\mathrm{G}$ Lagrangian is
\be\label{eq:scetglagrangian1}
{\cal L}_{\mathrm{SCET_G}}(\xi_n,A_n,A_\mathrm{G}) = {\cal L}_{\mathrm{SCET}}(\xi_n,A_n)+ {\cal L}_\mathrm{G}(\xi_n,A_n,A_\mathrm{G}) \, ,
\ee
where ${\cal L}_{\mathrm{SCET}}(\xi_n,A_n)$ is the vacuum SCET Lagrangian, with $\xi_n$ and $A_n$ the collinear quark and gluon fields, respectively. The second term ${\cal L}_\mathrm{G}(\xi_n,A_n,A_\mathrm{G})$ is given by
\be\label{eq:scetglagrangian2}
 {\cal L}_\mathrm{G}(\xi_n,A_n,A_\mathrm{G}) = g\sum_{\tilde p,\tilde p'}e^{-i(\tilde p-\tilde p')\cdot x}\left(\bar\xi_{n,p'}T^a\f{\slashed{\bar n}}{2}\xi_{n,p}-if^{abc}A_{n,p'}^{\lambda c}A_{n,p}^{\nu,b}g_{\nu\lambda}^\perp\bar n\cdot p\right) n\cdot A_\mathrm{G}^a\,,
\ee
which contains the interactions between the Glauber gluons $A_\mathrm{G}$ and the collinear quarks and gluons traversing the QCD medium. This result corresponds to the static source as described in more detail in~\cite{Ovanesyan:2011xy}. Moreover, the result presented here for ${\cal L}_{\mathrm{SCET_G}}$ corresponds to the so-called hybrid gauge, where a different gauge is chosen for the collinear gluons (light-cone gauge) than for the Glauber gluons (covariant $R_\xi$ gauge). This is a valid gauge choice as it was shown in~\cite{Ovanesyan:2011xy} since both sectors of the effective theory SCET$_\mathrm{G}$ are separately gauge invariant. This gauge choice simplifies the calculations of the in-medium splitting functions considerably as presented in the next section. In the hybrid gauge both the collinear Wilson line as well as the transverse gauge link which appear in the effective field theory reduce to unity. The corresponding Feynman rules for the interaction of collinear massless quarks and gluons with the Glauber modes can be obtained directly from Eq.~(\ref{eq:scetglagrangian2}). 

We now present the extension of the above effective field theory SCET$_\mathrm{G}$ by including finite quark masses. The vacuum SCET Lagrangian involving finite quark masses was first derived in~\cite{Leibovich:2003jd}. As mentioned above, the corresponding effective field theory is typically denoted by SCET$_{\mathrm{M}}$. In this section, we recall its derivation and, in addition, we include Glauber modes that describe the interaction with the QCD medium. The resulting new effective field theory that involves both massive quarks and Glauber modes describing the interaction with the medium is denoted by SCET$_\mathrm{M,G}$. We only focus on the collinear quark sector of the Lagrangian density for which we now take into account finite quark masses. We start from the standard QCD Lagrangian
\be\label{eq:LQCD}
{\cal L}_{\mathrm{QCD}}=\bar\psi(i\slashed{D}-m)\psi \, ,
\ee
where the covariant derivative is given by $iD^\mu = \partial^\mu + gA^\mu$. Here the gauge field $A^\mu$ consists of three contributions
\be
A^\mu = A^\mu_c+A^\mu_s+A^\mu_\mathrm{G} \, ,
\ee
where $A^\mu_{c,s,\mathrm{G}}$ are the collinear, the soft, and the Glauber gluon gauge fields, respectively. The collinear and soft fields scale like the corresponding collinear and soft momenta as described above. The Glauber gluon has the momentum scaling as $p^\mu\sim p^+(\lambda^2,\lambda^2,\lambda)$. However, the scaling of the corresponding Glauber gluon field $A_\mathrm{G}$ is different and needs to be derived by expressing the Glauber gluon field in terms of the QCD current of the source and the gluon propagator. By working out the scaling of every term in the resulting expression, the scaling for $A_\mathrm{G}$ can be obtained which eventually also depends on the gauge choice, see~\cite{Idilbi:2008vm,Ovanesyan:2011xy} for more details. In the hybrid gauge we use the covariant $R_\xi$ gauge for the medium Glauber gluons. In this case the corresponding Glauber gluon field scales as $A_\mathrm{G}\sim p^+(\lambda^2,\lambda^2,\lambda^3)$ as it was derived in~\cite{Ovanesyan:2011xy}. We start by substituting Eqs.~(\ref{eq:psi1}) and~(\ref{eq:psi2}) into the QCD Lagrangian in Eq.~\eqref{eq:LQCD}, and we find
\ba\label{eq:Lfull}
{\cal L} & = & \sum_{\tilde p,\tilde p'}e^{-i(\tilde p-\tilde p')\cdot x}\left[\bar\xi_{n,p'}\f{\slashed{\bar n}}{2} in\cdot D\,\xi_{n,p} + \bar\xi_{\bar n,p'}\f{\slashed{n}}{2}(\bar{\cal P}+i\bar n\cdot D)\xi_{\bar n,p} \right. \nn \\
&&\hspace*{2.5cm}+ \bar\xi_{n,p'}(\slashed{\mathcal{P}}_\perp+i\slashed{D}_\perp-m)\xi_{\bar n,p}+\bar\xi_{\bar n,p'}(\slashed{\mathcal{P}}_\perp+i\slashed{D}_\perp-m)\xi_{n,p} \Big] \, .
\ea
We can now integrate out the small component of the collinear quark field $\xi_{\bar n,p}$ by making use of the equation of motion
\be
(\bar{\cal P}+i\bar n\cdot D) \xi_{\bar n,p}=(\slashed{\mathcal{P}}_\perp+i\slashed{D}_\perp+m)\f{\slashed{\bar n}}{2}\xi_{n,p}\, .
\ee
With this relation, we obtain from Eq.~(\ref{eq:Lfull}) the following result for the leading-order (in $\lambda$) SCET$_{\mathrm{M,G}}$ Lagrangian density
\be\label{eq:L1}
{\cal L}_ 0 = \sum_{\tilde p,\tilde p',\tilde q}e^{-ix\cdot{\cal P}}\, \bar\xi_{n,p'}\left[in\cdot D+(\slashed{\mathcal{P}}_\perp+g\slashed{A}_{n,q}^\perp)W_n\f{1}{\bar{\cal P}}W_n^\dagger(\slashed{\mathcal{P}}_\perp+g\slashed{A}_{n,q'}^\perp) \right]\f{\slashed{\bar n}}{2}\xi_{n,p} + {\cal L}_m \, ,
\ee
where we introduced the label momentum operator also in the exponential for notational convenience and we have
\be
A_c^\mu=\sum_{\tilde q}e^{-i\tilde q\cdot x}A_{n,q}^\mu\, .
\ee
The mass-dependent terms ${\cal L}_m$ of the leading-order Lagrangian ${\cal L}_0$ in Eq.~(\ref{eq:L1}) are given by
\bea\label{eq:L2}
{\cal L}_m= \sum_{\tilde p,\tilde p',\tilde q}e^{-ix\cdot{\cal P}}\,  \left[m\, \bar\xi_{n,p'}\left[(\slashed{\mathcal{P}}_\perp+g\slashed{A}_{n,q}^\perp),W_n\f{1}{\bar{\cal P}}W_n^\dagger \right]\f{\slashed{\bar n}}{2}\xi_{n,p}-m^2\, \bar\xi_{n,p'}W_n\f{1}{\bar{\cal P}}W_n^\dagger\f{\slashed{\bar n}}{2}\xi_{n,p}\right] \, .
\eea
Whether ${\cal L}_m$ contributes at leading-order in $\lambda$ still depends on the scaling of $m$ as discussed further in the next section. Note that the factor $in\cdot D$ in Eq.~(\ref{eq:L1}) contains the collinear, the soft, and the Glauber gluon field
\be\label{eq:inD}
in\cdot D=in\cdot\partial+g\, n\cdot A_n+g\, n\cdot A_s+g\, n\cdot A_\mathrm{G} \, .
\ee
The last term here gives the interaction vertex of a collinear quark with the medium off-shell Glauber gluons. It is exactly the same as in the massless case, see Eq.~(\ref{eq:scetglagrangian2}) above. As it turns out, there is no modification for this vertex due to the finite quark mass when we work in the hybrid gauge. As mentioned above, the Glauber gluon field scales as $A_\mathrm{G}\sim p^+(\lambda^2,\lambda^2,\lambda^3)$. Following the usual power counting arguments, the only situation where we obtain such an interaction term is from Eq.~(\ref{eq:inD}). All other possible contributions are power corrections in $\lambda$ to the leading-order Lagrangian. In particular, there is no Glauber gluon term in ${\cal L}_m$ to leading-order in $\lambda$. In other words, the massive part of the vacuum SCET Lagrangian is not modified at leading-order when including medium interactions. 

To summarize, we find that the Feynman rules that describe the interaction with the medium remain the same as in the massless case~\cite{Ovanesyan:2011xy}. This statement holds in the hybrid gauge for a static source, which is the relevant case for all practical purposes considered in this work. All Feynman rules derived from the massive SCET$_\mathrm{M}$ Lagrangian in the vacuum also remain the same. In the next sections, we are going to make use of them when calculating the massive splitting functions in vacuum as well as in the QCD medium. Since SCET$_{\mathrm{M,G}}$ is a direct combination of SCET$_\mathrm{M}$ in the vacuum and SCET$_{\mathrm{G}}$ for the medium interactions, we do not repeat the relevant Feynman rules here but instead refer the reader to~\cite{Leibovich:2003jd,Ovanesyan:2011xy}.

\subsection{Massive splitting functions in the vacuum derived from SCET$_{\mathrm{M}}$}

In this section, we derive the massive splitting functions in vacuum for the channels $Q\to Qg$, $Q\to gQ$, and $g\to Q\bar Q$, using SCET$_\mathrm{M}$. The same results obtained in the standard perturbative QCD can be found in~\cite{Catani:2000ef}, where the authors derived their results using the so-called ``quasi-collinear'' limit. 
Such a limit is an extension of the ``collinear limit'' used for deriving massless splitting functions, where the on-shell quark masses are kept of the same order as the invariant mass of the two final state partons. See~\cite{Catani:2000ef} for more details. Using SCET$_\mathrm{M}$, this limit has already been taken into account at the level of the SCET$_\mathrm{M}$ Lagrangian when the mass is taken to scale as $m\sim p^+\lambda$. Therefore, the massive splitting functions can be obtained directly without having to make any further approximations.

\subsubsection{$Q\to Qg$}

\begin{figure*}[htb]
\centering
\includegraphics[width=12cm,clip]{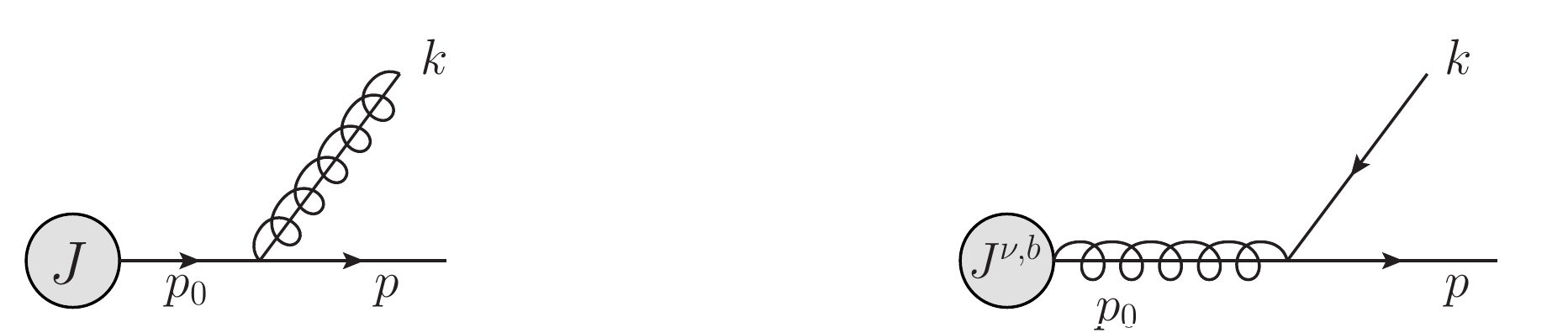}
\vspace*{-.1cm}
\caption{\label{fig1}  Feynman diagrams for the two splitting processes involving massive quarks $Q\to Q g$ (left) and $g\to Q\bar Q$ (right). The off-diagonal splitting process $Q\to gQ$ can be obtained from the result for $Q\to Qg$ via crossing. $J$ and $J^{\nu,b}$ represents the remaining amplitude that produces the incident highly energetic parent quark or gluon respectively.}
\end{figure*}

We adopt the notation introduced in~\cite{Ovanesyan:2011xy,Ovanesyan:2011kn}. We first consider the splitting process $Q(p_0)\to Q(p)+g(k)$, where we labeled the four momenta of the involved partons. The splitting process is illustrated in Fig.~\ref{fig1} (left). Throughout this paper, we adopt the convention to label a heavy quark of mass $m$ by $Q$ whereas a massless quark is denoted by $q$. As mentioned above, we work in the hybrid gauge for the in-medium splitting functions. Therefore, we adopt the light-cone gauge also for the vacuum calculation. Using the on-shell conditions $p^2=m^2$ and $k^2=0$ for the two outgoing partons as well as momentum conservation, we can parametrize the involved momenta as follows
\ba\label{eq:momenta1}
p_0 & = & \left[ p_0^+,\f{\vect{k}_\perp^2+xm^2}{x(1-x)p_0^+},\vect{0}\right] \, ,\nn\\
k & = & \left[xp_0^+,\f{\vect{k}_\perp^2}{xp_0^+},\vect{k}_\perp\right] \, ,\nn \\
p & = & \left[(1-x)p_0^+,\f{\vect{k}_\perp^2+m^2}{(1-x)p_0^+},-\vect{k}_\perp \right] \, .
\ea
Here, we choose to work in the frame in which the parent parton has no transverse momentum,  $x=k^+/p_0^+$ is the momentum fraction taken away by the emitted gluon, and $\vc{k}_\perp$ is the gluon momentum transverse to the parent parton momentum. 

The amplitude ${\cal A}^{\mathrm{vac}}_{Q\to Qg}$ for the splitting process can be written as
\be\label{eq:amplitude1}
{\cal A}^{\mathrm{vac}}_{Q\to Qg}=\bar\xi_{n,p} \, ig\, T^a R^\mu(p,k,m) \, \f{\slashed{\bar n}}{2}\, \f{i\slashed{n}}{2}\f{\bar n\cdot (p+k)}{(p+k)^2-m^2}\, \varepsilon_\mu(k)\, J \, ,
\ee
where $J$ stands for the remaining amplitude producing the massive parent quark $Q$ that undergoes the splitting process, see Fig.~\ref{fig1} (left). The factor $R^\mu(p,k,m)$ is associated with the splitting vertex. Using the Feynman rules for SCET$_{\mathrm{M}}$, we find
\bea
R^\mu(p,k,m)  = & n^\mu+\f{\gamma_\perp^\mu(\slashed{p}_\perp+\slashed{k}_\perp)}{\bar n\cdot (p+k)}+\f{\slashed{p}_\perp\gamma_\perp^\mu}{\bar n\cdot p}-\f{\slashed{p}_\perp(\slashed{p}+\slashed{k})_\perp}{\bar n\cdot p\,\bar n\cdot(p+k)}\bar n^\mu \nn \\
& + \f{m}{\bar n \cdot (p+k)\,\bar n\cdot p}\left[\gamma_\perp^\mu(\bar n\cdot p-\bar n\cdot (p+k))+\bar n^\mu((\slashed{p}+\slashed{k})_\perp-\slashed{p}_\perp+m) \right] \,. 
\eea
The gluon polarization vector $\varepsilon^\mu(k)$ in Eq.~(\ref{eq:amplitude1}) can be written as~\cite{Gyulassy:2000er,Kang:2012vm}
\be\label{eq:eps}
\varepsilon^\mu(k)=\left[0,\f{2\vect{\varepsilon}^i_{\perp}\cdot \vect{k}_\perp}{k^+}, \vect{\varepsilon}^i_{\perp} \right]
\ee
satisfying both $k\cdot\varepsilon(k)=0$ and $\bar n\cdot\varepsilon(k)=0$. Using the parametrization of the momenta as defined in~(\ref{eq:momenta1}), we can express the factor resulting from the massive quark propagator as
\be
\f{\bar n\cdot(p+k)}{(p+k)^2-m^2}=\frac{x(1-x)p_0^+}{\vect{k}_\perp^2+x^2m^2}\, .
\ee
Furthermore, we can write the product $R(p,k,m) \cdot \varepsilon(k)$ as
\ba\label{eq:Reps}
R(p,k,m)\cdot\varepsilon(k) & = & \f{\vect{\varepsilon}^i_\perp}{x(1-x)p_0^+} \left[2\vect{k}_\perp^i+x \gamma_\perp^i\gamma_\perp^j \vect{k}_\perp^j +\gamma_\perp^i mx^2\right] \nn \\ 
& \equiv &  \f{\vect{\varepsilon}^i_\perp}{x(1-x)p_0^+} (\Gamma^i + \Gamma_m^i) \, .
\ea
Here, we separated the resulting three terms into two pieces where $\Gamma_m^i$ contains only the third term proportional to the mass $m$ in the first line of Eq.~\eqref{eq:Reps}, since $\Gamma_m^i$ has an odd number of $\gamma_\perp$ matrices. In summary, we can write the amplitude for the splitting process as
\be
{\cal A}^{\mathrm{vac}}_{Q\to Qg}=-gT^a \, \bar\xi_{n,p}\,(\Gamma^i + \Gamma_m^i)\, \vect{\varepsilon}_\perp^i\, \f{x(1-x)p_0^+}{\vect{k}_\perp^2+x^2m^2}\, J.
\ee

To proceed we square the amplitude ${\cal A}^{\mathrm{vac}}_{Q\to Qg}$ and average over spin and color configurations of the initial quark:
\be\label{eq:qqqqg}
\f{1}{2N_c}|{\cal A}^{\mathrm{vac}}_{Q\to Qg}|^2 = g^2\,C_F\,\mathrm{Tr}\left[\f{\slashed{n}}{2}\bar n\cdot p\, J\bar J\gamma^0(\Gamma^i+\Gamma_m^i)^\dagger\gamma^0(\Gamma^i-\Gamma^i_m)\right]\f{x^2(1-x)^2(p_0^+)^2}{(\vect{k}_\perp^2+x^2m^2)^2} \,,
\ee
where we used
\be
\sum\vect{\varepsilon}_\perp^i\vect{\varepsilon}_\perp^{i'}=\delta^{ii'}\, , \quad \sum \xi_{n,p}\,\bar \xi_{n,p} = \f{\slashed{n}}{2}\bar n\cdot p \, .
\ee
Note that the second relation is the same for massless and massive collinear quarks. We can now write the expression in Eq.~(\ref{eq:qqqqg}) involving the $\Gamma^i$ matrices inside the trace as
\be
\gamma^0(\Gamma^i+\Gamma_m^i)^\dagger\gamma^0(\Gamma^i-\Gamma^i_m) = \f{1}{x^2(1-x)^2(p_0^+)^2}\left[4(1-x+x^2/2)\vect{k}_\perp^2+2x^4m^2\right]\mathbb{I}_{\mathrm{Dirac}}\, \mathbb{I}_{\mathrm{color}},
\ee
which is a scalar in both Dirac and color space. Here we used
\be
\gamma_\perp^i\gamma_\perp^j = -\delta^{ij} - i\varepsilon^{ij3}\Sigma^3,\;\mathrm{where}\quad \Sigma^3=\left(\begin{array}{cc}
\sigma^3 & 0 \\
0 & \sigma^3 \\
\end{array} \right)\, ,
\ee
and $(\Sigma^3)^\dagger=\Sigma^3$ and $(\Sigma^3)^2=\mathbb{I}$. Hence, we may now write the splitting amplitude squared in a factorized form
\bea\label{eq:AQ}
\f{1}{2N_c}|{\cal A}^{\mathrm{vac}}_{Q\to Qg}|^2  = & \f{1}{2} \mathrm{Tr}\left[\f{\slashed{n}}{2}\bar n\cdot p_0\, J\bar J\right]\times g^2 C_F\f{4x(1-x)}{\vect{k}_\perp^2+x^2m^2}\left[\f{1-x+x^2/2}{x}-\f{2x(1-x)m^2}{\vect{k}_\perp^2+x^2m^2} \right] \nn\\
\equiv & |{\cal A}^{\mathrm{vac}}_{Q}|^2 \times |{\cal A}^{\mathrm{vac}}_{Q\to Qg}|^2 \, ,
\eea
where the first factor is the leading-order amplitude squared, {\it i.e.} without any emission. In order to obtain the correct normalization for the splitting function, we need to factor out $|{\cal A}^{\mathrm{vac}}_{Q}|^2$. Thus the splitting function for the process $Q\to Qg$ will be given by the second factor on the right-hand side of Eq.~\eqref{eq:AQ}.

We still need to take into account the phase space for the quark (momentum $p$) and the gluon (momentum $k$) in the final state. Following for example~\cite{Baumgart:2010qf}, we write the corresponding 2-particle phase space as
\ba
\f{d^3p}{2p^0}\f{d^3k}{2k^0} & = & \f{dp^+d^2\vect{p}_\perp}{2p^+} \f{dk^+d^2\vect{k}_\perp}{2k^+}\,d^4p_0\,\delta^{(4)}(p_0-p-k)\nn \\
& = & \f{dp_0^+d^2\vect{p}_{0\perp}}{2p_0^+}\times \f{dx\, d^2\vect{k}_\perp}{2x(1-x)}\, .
\ea
Here, the first factor is the leading-order phase space (momentum $p_0$) which needs to be factored out together with the leading-order amplitude squared $|{\cal A}^{\mathrm{vac}}_{Q}|^2$ in Eq.~\eqref{eq:AQ}, in order to obtain the correct normalization for the splitting function. Eventually, the $Q\to Q g$ massive splitting function in the vacuum is given by
\be\label{eq:massiveQQg}
\left(\f{dN^{\mathrm{vac}}}{dxd^2\vect{k}_\perp} \right)_{Q\to Qg} = C_F\f{\as}{\pi^2}\f{1}{\vect{k}_\perp^2+x^2m^2}\left[\f{1-x+x^2/2}{x}-\f{x(1-x)m^2}{\vect{k}_\perp^2+x^2m^2} \right] \, ,
\ee
which reduces to the massless splitting function derived analogously in~\cite{Ovanesyan:2011xy,Ovanesyan:2011kn} when taking $m\to 0$. Different to the massless case, the dependence on $x$ and $\vc{k}_\perp$ does not factorize anymore. 

Likewise, we can derive the splitting kernel for the process $Q\to gQ$. Being a crossed process of $Q\to Qg$, the splitting function for $Q\to gQ$ can be obtained from Eq.~\eqref{eq:massiveQQg} by substituting $x\to 1-x$. We now turn to the process $g\to Q\bar Q$. 

\subsubsection{$g\to Q\bar Q$}

Next, we consider the splitting process where a gluon splits into a massive quark anti-quark pair: $g(p_0)\to Q(p)+\bar Q(k)$, as shown in Fig.~\ref{fig1} (right). Analogous to Eq.~(\ref{eq:momenta1}), we may now write the momenta of the involved partons as
\ba
p_0 & = & \left[ p_0^+,\f{\vect{k}_\perp^2+m^2}{x(1-x)p_0^+},\vect{0}\right] \, ,\nn\\
k & = & \left[xp_0^+,\f{\vect{k}_\perp^2+m^2}{xp_0^+},\vect{k}_\perp\right] \, ,\nn \\
p & = & \left[(1-x)p_0^+,\f{\vect{k}_\perp^2+m^2}{(1-x)p_0^+},-\vect{k}_\perp \right] \, .
\ea
The amplitude for the splitting process $g\to Q\bar Q$ can be written as
\be\label{eq:ampgQQ}
{\cal A}^{\mathrm{vac}}_{g\to Q\bar Q}=\bar \xi_{n,p}\, igT^a\,{R'}^\mu(p,k,m)\f{\slashed{\bar n}}{2}\,\xi_{n,k}\f{-i\delta^{ab}N_{\mu\nu}}{(p+k)^2}\, J^{\nu,b} \, .
\ee
The function $J^{\nu,b}$ represents the remaining amplitude that produces the parent gluon, with Lorentz and color indices $\nu$ and $b$ respectively. We assume again the physical polarization for the parent gluon, and thus $J^{\nu,b}$ satisfies $\bar n\cdot J=(p+k)\cdot J=0$, cf. Eq.~(\ref{eq:eps}). In addition, we have
\ba
N_{\mu\nu} & = & \left(g_{\mu\nu}-\f{\bar n_{\mu}(p+k)_\nu+\bar n_\nu (p+k)_\mu}{\bar n\cdot(p+k)}\right) \,, \\
{R'}^\mu (p,k,m) & = & n^\mu+\f{\gamma_\perp^\mu \slashed{k}_\perp}{\bar n\cdot k}+\f{\slashed{p}_\perp\gamma_\perp^\mu}{\bar n\cdot p}-\f{\slashed{p}_\perp\slashed{k}_\perp}{\bar n\cdot p\,\bar n\cdot k}\bar n^\mu \nn \\
&& - \f{m}{\bar n \cdot k\,\bar n\cdot p}\left[\gamma_\perp^\mu(\bar n\cdot p+\bar n\cdot k)+\bar n^\mu(-\slashed{k}_\perp-\slashed{p}_\perp+m) \right] \, .
\ea
Note that the factor ${R'}^\mu(p,k,m)$ associated with the splitting vertex depends on the direction of the momentum flow. Analogous to Eq.~(\ref{eq:Reps}), we can rewrite the combination ${R'}^\mu(p,k,m) N_{\mu\nu} J^{\nu,b}$ as
\ba
{R'}^\mu(p,k,m) N_{\mu\nu} J^{\nu,b} & = & \f{\vect{J}_\perp^{i,b}}{x(1-x)p_0^+}\left(2x\vect{k}_\perp^i+\gamma_\perp^i\gamma_\perp^j\vect{k}_\perp^j+m\gamma_\perp^i\right) \nn \\
& \equiv & \f{\vect{J}_\perp^{i,b}}{x(1-x)p_0^+} (\Gamma^i+\Gamma^i_m)\, .
\ea
In the second line, we separated again the $m$-dependent part from the rest as it contains only one $\gamma_\perp$ matrix. Squaring the amplitude ${\cal A}^{\mathrm{vac}}_{g\to Q\bar Q}$ in Eq.~(\ref{eq:ampgQQ}) and averaging over gluon polarizations and colors, we obtain the following result
\be\label{eq:sq2}
\f{1}{2(N_c^2-1)}|{\cal A}^{\mathrm{vac}}_{g\to Q\bar Q}|^2 = \f{T_R}{2}\mathrm{Tr}\left[\f{\slashed{n}}{2}\bar n\cdot k\,\f{\slashed{\bar n}}{2}\f{\slashed{n}}{2}\bar n\cdot p\f{\slashed{\bar n}}{2}\gamma^0 (\Gamma^i+\Gamma^i_m)^\dagger\gamma^0 (\Gamma^i-\Gamma^i_m)\right] \f{x^2(1-x)^2}{(\vect{k}_\perp^2+m^2)^2}\, .
\ee
Here we used $\vect{J}_\perp^{i,b} \vect{J}_\perp^{i',b'}=\delta^{ii'}\delta^{bb'}$ which is the appropriate relation for the calculation of the spin averaged splitting function, see e.g.~\cite{Catani:1998nv}. We further evaluate the expressions containing the $\Gamma^i$'s and find
\bea
\gamma^0(\Gamma^i+\Gamma_m^i)^\dagger\gamma^0(\Gamma^i-\Gamma_m^i)  = & \nn \\
&\hspace*{-4.5cm} \f{2}{x^2(1-x)^2(p_0^+)^2} \left[(x^2+(1-x)^2)(\vect{k}_\perp^2+m^2)+2x(1-x)m^2 \right] \mathbb{I}_{\mathrm{Dirac}}\, \mathbb{I}_{\mathrm{color}} \, .
\eea
We continue by evaluating the remaining trace part in Eq.~(\ref{eq:sq2}) and by including the appropriate phase space factors as before. After taking into account the normalization to the leading-order amplitude squared, we obtain the following result for the $g\to Q\bar Q$ splitting function in the vacuum
\be\label{eq:massivegQQbar}
\left(\f{dN^{\mathrm{vac}}}{dxd^2\vect{k}_\perp} \right)_{g\to Q\bar Q} = T_R \f{\as}{2\pi^2}\f{1}{\vect{k}_\perp^2+m^2}\left[x^2+(1-x)^2+\f{2x(1-x)m^2}{\vect{k}_\perp^2+m^2} \right],
\ee
which reduces to the massless splitting function derived analogously in~\cite{Ovanesyan:2011kn}. Again, the dependence on $x$ and $\vect{k}_\perp$ does not factorize as in the massless case. 

\subsection{Massive splitting functions in the medium derived from SCET$_{\mathrm{M,G}}$}

\begin{figure*}[t!]
\centering
\includegraphics[width=\textwidth,trim=0cm 11cm 0cm 11cm]{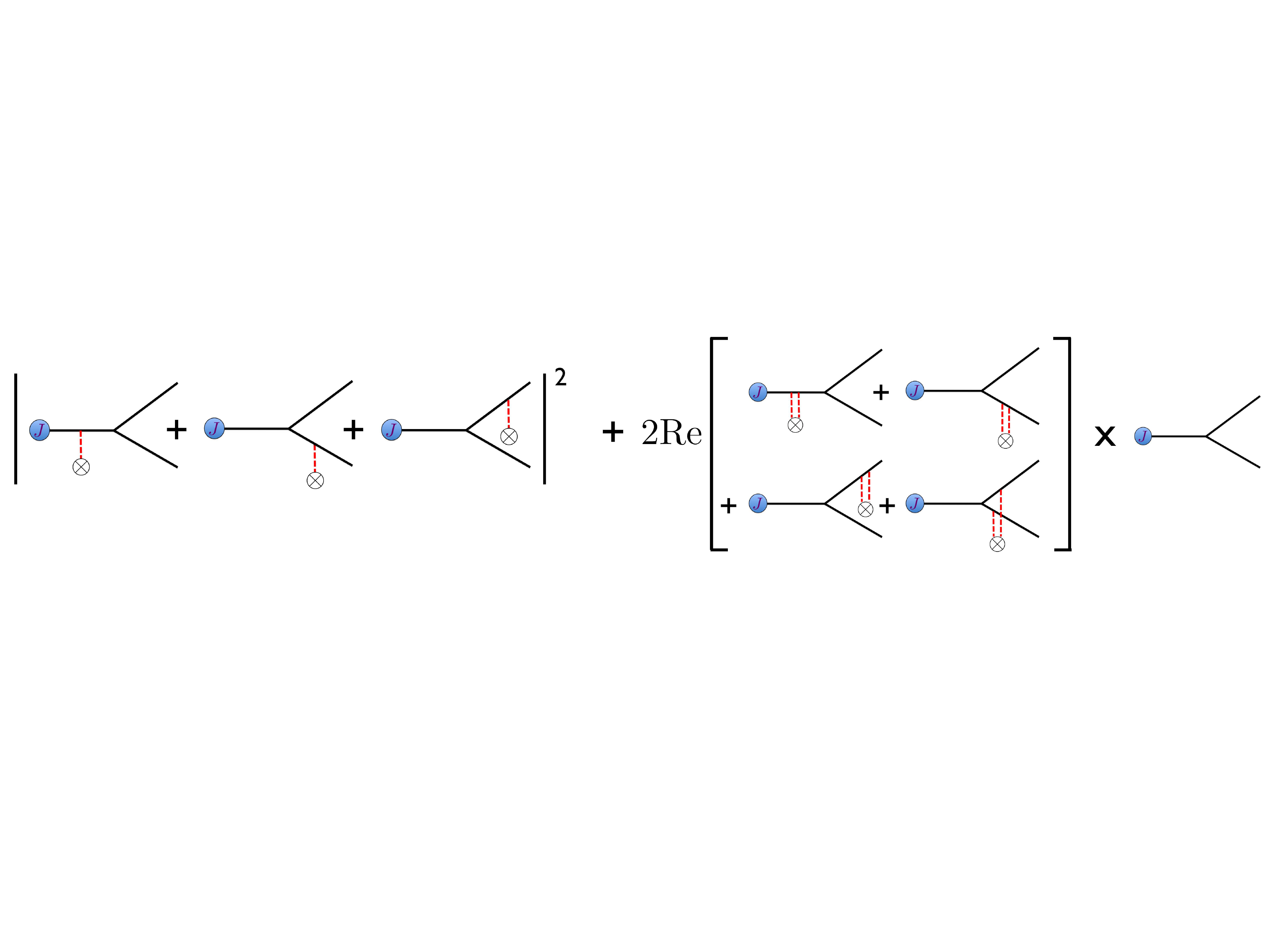}
\vspace*{-.1cm}
\caption{\label{fig:in-medium-diagrams}  Single- and double-Born Feynman diagrams that contribute to the massive in-medium splitting functions to first order in opacity. The topology is the same for all three splitting processes $Q\to Qg$, $Q\to gQ$ and $g\to Q\bar Q$. Figure adapted from~\cite{Ovanesyan:2011kn}.}
\end{figure*}

The calculation of the massive in-medium splitting functions follows roughly the same steps as in the massless case~\cite{Ovanesyan:2011xy}. For completeness, we outline the basic steps of the calculation. For every Glauber interaction of the energetic collinear parton with the $i$'th scattering center, we need to integrate over the Glauber gluon momentum. We denote the corresponding integral by $d\Phi_i$. Following~\cite{Ovanesyan:2011xy}, we introduce the following notation
\be
d\Phi_i=\f{d^4q_i}{(2\pi)^4}e^{iq_i\delta x_i}v(q_i),\qquad d\vc{\Phi}_{i\perp}=\f{d^2\vc{q}_{i\perp}}{(2\pi)^2}e^{-i\vc{q}_{i\perp}\delta\vc{x}_i}\tilde v(\vc{q}_{i\perp}) \, ,
\ee
where $q_i$ is the momentum exchanged between the incident parton and the QCD medium (through the Glauber gluon exchange) and $\vc{q}_{i\perp}$ is its transverse component. Moreover, we have $\delta x_i=x_i-x_0$, where $x_0$ is the space-time position where the initial energetic parton was created and $x_i$ is the position of the $i$'th interaction with the quasi-particles of the medium. The transverse component of $\delta x_i$ is denoted by $\delta\vc{x}_i$. The functions $v(q_i)$ and $\tilde v(\vc{q}_{i\perp})$ are related to the elastic scattering cross section. We have $v(q_i)=2\pi \delta(q^0_i)\tilde v(\vc{q}_{i\perp})$ and to lowest order for a Yukawa-screened potential
\be\label{eq:sigmael}
\f{d\sigma_{el}}{d^2\vc{q}_\perp}(R,T)=\f{C_2(R)C_2(T)}{8}\f{|\tilde v(\vc{q}_\perp)|^2}{(2\pi)^2} = \f{C_2(R)C_2(T)}{8} \f{4\as^2}{(\vc{q}_\perp^2+\mu^2)^2}\, ,
\ee
where $C_2(R)$ and $C_2(T)$ denote the quadratic Casimir invariants in the representation of the incident highly energetic parton and the target (source) respectively. See~\cite{Ovanesyan:2011xy} for more details. The delta function in $v(q_i)$ makes the $q^+$ integral trivial. We can now write $d\Phi_i$ as
\be
d\Phi_i=d\vc{\Phi}_{i\perp}\f{dq_i^-}{2\pi} e^{iq_i^-\delta z_i} \, ,
\ee
where $\delta z_i$ is the distance along the $z$-axis between the scattering center $i$ and the point where initial energetic parton was created. Remaining integrals over $\vc{q}_{\perp}$ will eventually be performed numerically at the end using a realistic model for the medium. However, the $q^-_i$ integrals still need to be performed analytically. The corresponding longitudinal integrals can be evaluated in terms of contour integrals in the complex plane. The results including non-vanishing mass terms can be obtained analogously to the techniques outlined in~\cite{Ovanesyan:2011xy} for the massless case. Formally, the amplitudes for the three in-medium splitting processes, can be written as
\ba
{\cal A}^{\mathrm{med}}_{Q\to ab} & = &  \braket{a(p)b(k)|T\bar\chi_n(x_0)e^{i\int d^4x\, {\cal L}_{\mathrm{SCET_{M,G}}}}|Q(p_0)}\, ,\\
{\cal A}^{\mathrm{med}}_{g\to Q\bar Q} & = & \braket{Q(p)\bar Q(k)|T {\cal B}^\mu_{n\perp}(x_0)e^{i\int d^4x\, {\cal L}_{\mathrm{SCET_{M,G}}}}|g(p_0)} \, ,
\ea
where we use the same labelling of the involved parton momenta as for the vacuum case in the previous section. Here, $Q\to ab$ corresponds to either the diagonal splitting $Q\to Qg$ or the off-diagonal case $Q\to gQ$. The gauge invariant quark and gluon fields $\bar\chi_n$ and ${\cal B}_{n\perp}^\mu$ were defined in~(\ref{eq:chiB}) and ${\cal L}_{\mathrm{SCET_{M,G}}}$ was derived in section~\ref{sec:SCETMG} above. In order to obtain the in-medium splitting functions to first order in opacity, we need to take into account both single- and double-Born diagrams for every scattering center $i$, see~\cite{Gyulassy:2000er}. We denote the corresponding single- and double Born amplitudes by ${\cal A}^{\mathrm{med}}_{\mathrm{SB}}$ and ${\cal A}^{\mathrm{med}}_{\mathrm{DB}}$ respectively. The double-Born diagrams are evaluated in the ``contact limit'' where $\delta z_1=\delta z_2$. After squaring the sum of all amplitudes, we have to calculate schematically
\be\label{eq:sbdb}
|{\cal A}^{\mathrm{med}}_{\mathrm{SB}}|^2 + 2\,\mathfrak{Re}\left\{{\cal A}^{\mathrm{med}}_{\mathrm{DB}}\times{\cal A}^{\mathrm{vac}} \right\} \, ,
\ee
where ${\cal A}^{\mathrm{vac}}$ is the vacuum splitting amplitude without any interaction with the QCD medium. In Fig.~\ref{fig:in-medium-diagrams}, all the relevant diagrams are shown that correspond to Eq.~(\ref{eq:sbdb}). The topology is the same for all three massive splitting processes. On the left hand side of Fig.~\ref{fig:in-medium-diagrams}, the square of the three single-Born diagrams is shown. On the right hand side, all double-Born diagrams are shown that give a non-zero result in the contact limit. For all double-Born diagrams, the interference with the vacuum leading-order splitting diagram is shown. After adding and squaring the relevant amplitudes, we still need to perform the sum over all scattering centers $i=1,\ldots N$. Following~\cite{Ovanesyan:2011xy}, this sum can be turned into a continuous integral that gives a delta function, which in turn can be used to perform one of the remaining transverse momentum integrals. We would like to stress that to first order in opacity, we take into account the single- and double-Born diagrams shown in Fig.~\ref{fig:in-medium-diagrams} for every scattering center $i$. See~\cite{Ovanesyan:2011xy} for more calculational details.

Since the intermediate steps of the calculations are very similar to the massless case~\cite{Ovanesyan:2011xy}, we will skip them and present only the final results. On the other hand, for comparison and for later convenience, we also list the results for the massless in-medium splitting functions as calculated in~\cite{Ovanesyan:2011xy,Ovanesyan:2011kn}. We define the transverse momentum vectors
\begin{eqnarray}\label{eq:transversemomentumvectors}
\vc{A}_{\perp}=\vc{k}_{\perp},\,\, \vc{B}_{\perp}=\vc{k}_{\perp} + x \vc{q}_{\perp} , \,\,
\vc{C}_{\perp}=\vc{k}_{\perp} -  (1-x)\vc{q}_{\perp},\,\, \vc{D}_{\perp}=\vc{k}_{\perp}-\vc{q}_{\perp},
\end{eqnarray}
where again $x$ and $\vc{k}_\perp$ are the longitudinal momentum fraction and the transverse momentum of the emitted parton relative to the parent parton respectively. Furthermore, $\vc{q}_\perp$ is the transverse momentum introduced by the Glauber gluon exchange. In addition, we have the following phases
\bea\label{eq:phases}
&\Omega_1-\Omega_2=\frac{\vc{B}_{\perp}^2}{p_0^+ x(1-x)}, \,\Omega_1-\Omega_3=\frac{\vc{C}_{\perp}^2}{p_0^+x(1-x)}, \,\, \Omega_2-\Omega_3=\frac{\vc{C}_{\perp}^2-\vc{B}_{\perp}^2}{p_0^+x(1-x)}, \,\,\nonumber\\
& \Omega_4=\frac{\vc{A}_{\perp}^2}{p_0^+x(1-x)},\,\, \Omega_5=\frac{\vc{A}_{\perp}^2-\vc{D}_{\perp}^2}{p_0^+x(1-x)}\,.
\eea
We reproduce the light parton  in-medium splitting functions for reference and subsequent comparison. The result for $q\to qg$ is given by
\bea
 &\hspace*{-0.35cm}  \left( \frac{dN^{\mathrm{med}}}{ dxd^2\vc{k}_{\perp} }\right)_{q\rightarrow qg}  =  \frac{\alpha_s}{2\pi^2}
C_F  \frac{1+(1-x)^2}{x}  
\int \frac{d\Delta z}{\lambda_g(z)}  
\int d^2{\vc q}_\perp  \frac{1}{\sigma_{el}} \frac{d\sigma_{el}^{\; {\mathrm{med}}}}{d^2 {\vc q}_\perp} \; 
 \Bigg[  \frac{\vc{B}_{\perp}}{\vc{B}_{\perp}^2} \mcdot \left( \frac{\vc{B}_{\perp}}{\vc{B}_{\perp}^2}  -  \frac{\vc{C}_{\perp}}{\vc{C}_{\perp}^2}   \right)
   \nonumber \\
& \hspace*{-1cm} \qquad \qquad  
  \times \big( 1-\cos[(\Omega_1 -\Omega_2)\Delta z] \big)  + \frac{\vc{C}_{\perp}}{\vc{C}_{\perp}^2} \mcdot \left( 2 \frac{\vc{C}_{\perp}}{\vc{C}_{\perp}^2}   
-    \frac{\vc{A}_{\perp}}{\vc{A}_{\perp}^2} - \frac{\vc{B}_{\perp}}{\vc{B}_{\perp}^2}  \right) \big(1- \cos[(\Omega_1 -\Omega_3)\Delta z] \big) \nonumber \\  
&\hspace*{-1cm}
   \qquad \qquad    + \frac{\vc{B}_{\perp}}{\vc{B}_{\perp}^2} \mcdot \frac{\vc{C}_{\perp}}{\vc{C}_{\perp}^2} 
\big( 1 -  \cos[(\Omega_2 -\Omega_3)\Delta z] \big)  
+ \frac{\vc{A}_{\perp}}{\vc{A}_{\perp}^2} \mcdot \left( \frac{\vc{D}_{\perp}}{\vc{D}_{\perp}^2} - \frac{\vc{A}_{\perp}}{\vc{A}_{\perp}^2} \right) 
\big(1-\cos[\Omega_4\Delta z]\big)  \nonumber \\
& \hspace*{-1cm}\qquad \qquad  -\frac{\vc{A}_{\perp}}{\vc{A}_{\perp}^2} \mcdot \frac{\vc{D}_{\perp}}{\vc{D}_{\perp}^2}\big(1-\cos[\Omega_5\Delta z]\big)   
+  \frac{1}{N_c^2}  \frac{\vc{B}_{\perp}}{\vc{B}_{\perp}^2} \mcdot  \left( \frac{\vc{A}_{\perp}}{\vc{A}_{\perp}^2}  -   
\frac{\vc{B}_{\perp}}{\vc{B}_{\perp}^2}      \right)
\big( 1-\cos[(\Omega_1 -\Omega_2)\Delta z] \big)   \Bigg] \, .
\label{CohRadSX1} 
\eea
Here $\lambda_{g,q}(z)$ is the gluon (quark) mean free path in the medium. $(1/\sigma_{\mathrm{el}})d\sigma^{\mathrm{med}}_{\mathrm{el}}/d^2\vc{q}_\perp$ is the normalized in-medium elastic scattering cross section, see~(\ref{eq:sigmael}) above. We are left with a three dimensional integral over $\vc{q}_\perp$ and $\Delta z$ that need to be evaluated using a realistic model for the QCD medium as it is produced in heavy-ion collisions. The latter is an integral over the interactions with the medium quasi-particles $0<\Delta z<L$, where $L$ is the size of the medium. Note that the in-medium splitting function for $q\to gq$ can be obtained via crossing $x\leftrightarrow 1-x$. The $\vc{q}_\perp$ integral will be evaluated numerically and is subject to phase space constraints. The results for the in-medium splitting processes $g\to gg$ and $g\to q\bar q$ are given by
\begin{eqnarray}
&& \hspace*{-0cm}  \left( \frac{dN^{\mathrm{med}}}{ dxd^2\vc{k}_{\perp} }\right)_{ \left\{ \begin{array}{c}   g \rightarrow gg\\     g\rightarrow q\bar{q}  \end{array} \right\} } 
= 
 \left\{ \begin{array}{c}     \frac{\alpha_s}{2\pi^2} \, 2 C_A \left(\frac{x}{1-x}+\frac{1-x}{x}+x(1-x) \right)  \\[1ex]
                           \frac{\alpha_s}{2\pi^2}  T_R \left( x^2+(1-x)^2 \right)  \end{array} \right\}
\int {d\Delta z}   \left\{ \begin{array}{c}    \frac{1}{\lambda_g(z)} \\[1ex]  \frac{1}{\lambda_q(z)}    \end{array} \right\}
\int d^2{\vc q}_\perp  \;  \nonumber \\
&&\hspace*{-1cm} \qquad \qquad
\times  \frac{1}{\sigma_{el}} \frac{d\sigma_{el}^{\; {\mathrm{med}}}}{d^2 {\vc q}_\perp} \Bigg[  2\, \frac{\vc{B}_{\perp}}{\vc{B}_{\perp}^2} \mcdot \left(\frac{\vc{B}_{\perp}}{\vc{B}_{\perp}^2}-\frac{\vc{A}_{\perp}}{\vc{A}_{\perp}^2}\right) \big( 1-\cos[(\Omega_1 -\Omega_2)\Delta z]  \big)   
+2\, \frac{\vc{C}_{\perp}}{\vc{C}_{\perp}^2} \mcdot \left(\frac{\vc{C}_{\perp}}{\vc{C}_{\perp}^2}-\frac{\vc{A}_{\perp}}{\vc{A}_{\perp}^2}\right)
\nonumber \\
 &&\hspace*{-1cm}\qquad \qquad \times  \big( 1-\cos[(\Omega_1 -\Omega_3)\Delta z]  \big) +  \left\{ \begin{array}{c}   - \frac{1}{2}   \\[1ex] \frac{1}{N_c^2-1} \end{array} \right\}
\Bigg(2 \frac{\vc{B}_{\perp}}{\vc{B}_{\perp}^2}\mcdot\left(\frac{\vc{C}_{\perp}}{\vc{C}_{\perp}^2}-\frac{\vc{A}_{\perp}}{\vc{A}_{\perp}^2}\right)\big(1-\cos[(\Omega_1-\Omega_2)\Delta z ]\big)
\nonumber\\
 &&\hspace*{-1cm}\qquad \qquad  +2\,\frac{\vc{C}_{\perp}}{\vc{C}_{\perp}^2}\mcdot\left(\frac{\vc{B}_{\perp}}{\vc{B}_{\perp}^2}-\frac{\vc{A}_{\perp}}{\vc{A}_{\perp}^2}\right)\big(1-\cos[(\Omega_1-\Omega_3)\Delta z]\big)-2\,\frac{\vc{C}_{\perp}}{\vc{C}_{\perp}^2}\mcdot \frac{\vc{B}_{\perp}}{\vc{B}_{\perp}^2}\big(1-\cos[(\Omega_2-\Omega_3)\Delta z]\big)
\nonumber\\
 &&\hspace*{-1cm}\qquad \qquad +2\,\frac{\vc{A}_{\perp}}{\vc{A}_{\perp}^2}\mcdot\left(\frac{\vc{A}_{\perp}}{\vc{A}_{\perp}^2}-\frac{\vc{D}_{\perp}}{\vc{D}_{\perp}^2}\right)\big(1-\cos[\Omega_4\Delta z]\big)+2\,\frac{\vc{A}_{\perp}}{\vc{A}_{\perp}^2}\mcdot \frac{\vc{D}_{\perp}}{\vc{D}_{\perp}^2}\big(1-\cos[\Omega_5\Delta z]\big)\Bigg) \Bigg] \, .
\label{CohRadSX2} 
\end{eqnarray}
Note that for all four massless splitting functions, the $x$-dependent vacuum splitting function factors out. 

Next, we present our final results for the massive in-medium splitting functions $Q\to Qg$, $Q\to gQ$ and $g\to Q\bar Q$. All transverse momentum vectors defined in Eq.~(\ref{eq:transversemomentumvectors}) remain the same, so are the phase factors $\Omega_2-\Omega_3$ and $\Omega_5$ in Eq.~(\ref{eq:phases}). However, the remaining three phases in Eq.~(\ref{eq:phases}) are modified as follows
\be\label{eq:massivephases}
\Omega_1-\Omega_2=\frac{\vc{B}_{\perp}^2+\nu^2}{p_0^+ x(1-x)}, \,\Omega_1-\Omega_3=\frac{\vc{C}_{\perp}^2+\nu^2}{p_0^+x(1-x)}, \,\, \Omega_4=\frac{\vc{A}_{\perp}^2+\nu^2}{p_0^+x(1-x)}, \,\,\nonumber\\
\ee
where the variable $\nu$ is given by
\ba\label{eq:nu}
\label{eq:nu1} \nu & = & x\, m\;\qquad\quad\;\; (Q\to Qg)\, ,\\
\label{eq:nu2} \nu & = & (1-x)\, m\;\quad (Q\to gQ) \, , \\
\label{eq:nu3} \nu & = & m\;\quad\qquad\quad\; (g\to Q\bar Q)\, ,
\ea
for the three different massive splitting processes, respectively. The full in-medium splitting function for $Q\to Qg$ is given by
\begin{eqnarray}
 &&\hspace*{-0.35cm}  \left( \frac{dN^{\mathrm{med}}}{ dxd^2\vc{k}_{\perp} }\right)_{Q\rightarrow Qg}  =  \frac{\alpha_s}{2\pi^2}
C_F 
\int \frac{d\Delta z}{\lambda_g(z)}  
\int d^2{\vc q}_\perp  \frac{1}{\sigma_{el}} \frac{d\sigma_{el}^{\; {\mathrm{med}}}}{d^2 {\vc q}_\perp} \; 
\Bigg\{\left(\frac{1+(1-x)^2}{x} \right) \Bigg[  \frac{\vc{B}_{\perp}}{\vc{B}_{\perp}^2+\nu^2}  \nonumber \\
&& \hspace*{-1cm} \qquad
  \times  \left( \frac{\vc{B}_{\perp}}{\vc{B}_{\perp}^2+\nu^2}  -  \frac{\vc{C}_{\perp}}{\vc{C}_{\perp}^2+\nu^2}   \right) \big( 1-\cos[(\Omega_1 -\Omega_2)\Delta z] \big)  + \frac{\vc{C}_{\perp}}{\vc{C}_{\perp}^2+\nu^2} \mcdot \left( 2 \frac{\vc{C}_{\perp}}{\vc{C}_{\perp}^2+\nu^2}   
-    \frac{\vc{A}_{\perp}}{\vc{A}_{\perp}^2+\nu^2} \right. \nonumber \\  
&&\hspace*{-1cm}
   \qquad  \left. - \frac{\vc{B}_{\perp}}{\vc{B}_{\perp}^2+\nu^2}  \right) \big(1- \cos[(\Omega_1 -\Omega_3)\Delta z] \big) + \frac{\vc{B}_{\perp}}{\vc{B}_{\perp}^2+\nu^2} \mcdot \frac{\vc{C}_{\perp}}{\vc{C}_{\perp}^2+\nu^2} 
\big( 1 -  \cos[(\Omega_2 -\Omega_3)\Delta z] \big)  
 \nonumber \\
&& \hspace*{-1cm}\qquad + \frac{\vc{A}_{\perp}}{\vc{A}_{\perp}^2+\nu^2} \mcdot \left( \frac{\vc{D}_{\perp}}{\vc{D}_{\perp}^2+\nu^2} - \frac{\vc{A}_{\perp}}{\vc{A}_{\perp}^2+\nu^2} \right) 
\big(1-\cos[\Omega_4\Delta z]\big)   -\frac{\vc{A}_{\perp}}{\vc{A}_{\perp}^2+\nu^2} \mcdot \frac{\vc{D}_{\perp}}{\vc{D}_{\perp}^2+\nu^2}\big(1-\cos[\Omega_5\Delta z]\big)   \nn\\  
&&  \hspace*{-1cm}\qquad +  \frac{1}{N_c^2}  \frac{\vc{B}_{\perp}}{\vc{B}_{\perp}^2+\nu^2} \mcdot  \left( \frac{\vc{A}_{\perp}}{\vc{A}_{\perp}^2+\nu^2}  -  \frac{\vc{B}_{\perp}}{\vc{B}_{\perp}^2+\nu^2}      \right)
\big( 1-\cos[(\Omega_1 -\Omega_2)\Delta z] \big)   \Bigg] \nn \\
&&  \hspace*{-1cm}\qquad +\,x^3m^2 \Bigg[  \frac{1}{\vc{B}_{\perp}^2+\nu^2} \mcdot \left( \frac{1}{\vc{B}_{\perp}^2+\nu^2}  -  \frac{1}{\vc{C}_{\perp}^2+\nu^2}   \right) \big( 1-\cos[(\Omega_1 -\Omega_2)\Delta z] \big)  +\,\ldots \Bigg] \Bigg\}
\label{eq:fullmassive1} 
\end{eqnarray} 
where the ellipses denote analogous terms as in the first square bracket following the pattern as indicated. The variable $\nu$ for the process $Q\to Qg$ was defined in Eq.~(\ref{eq:nu1}), i.e. $\nu=x\,m$. The expressions in both square brackets have the same structure as the full massless in-medium splitting functions. The result for $Q\to gQ$ can be obtained via crossing. The splitting function $g\to Q\bar Q$ is given by
\begin{eqnarray}
&& \hspace*{-0.35cm}  \left( \frac{dN^{\mathrm{med}}}{ dxd^2\vc{k}_{\perp} }\right)_{g\to Q\bar Q} 
= \frac{\alpha_s}{2\pi^2}  T_R
\int {d\Delta z}    \frac{1}{\lambda_q(z)}
\int d^2{\vc q}_\perp  \frac{1}{\sigma_{el}} \frac{d\sigma_{el}^{\; {\mathrm{med}}}}{d^2 {\vc q}_\perp} \Bigg\{ \left( x^2+(1-x)^2 \right) \;  \nonumber \\
&&\hspace*{-1cm} \qquad 
\times \Bigg[  2\, \frac{\vc{B}_{\perp}}{\vc{B}_{\perp}^2+\nu^2} \mcdot \left(\frac{\vc{B}_{\perp}}{\vc{B}_{\perp}^2+\nu^2}-\frac{\vc{A}_{\perp}}{\vc{A}_{\perp}^2+\nu^2}\right) \big( 1-\cos[(\Omega_1 -\Omega_2)\Delta z]  \big)  \nonumber \\
 &&\hspace*{-1cm}\qquad +2\, \frac{\vc{C}_{\perp}}{\vc{C}_{\perp}^2+\nu^2} \mcdot \left(\frac{\vc{C}_{\perp}}{\vc{C}_{\perp}^2+\nu^2}-\frac{\vc{A}_{\perp}}{\vc{A}_{\perp}^2+\nu^2}\right)  \big( 1-\cos[(\Omega_1 -\Omega_3)\Delta z]  \big) +  \frac{1}{N_c^2-1}
\Bigg(2 \frac{\vc{B}_{\perp}}{\vc{B}_{\perp}^2+\nu^2} \nonumber\\
 &&\hspace*{-1cm}\qquad \times \left(\frac{\vc{C}_{\perp}}{\vc{C}_{\perp}^2+\nu^2}-\frac{\vc{A}_{\perp}}{\vc{A}_{\perp}^2+\nu^2}\right)\big(1-\cos[(\Omega_1-\Omega_2)\Delta z ]\big) +2\,\frac{\vc{C}_{\perp}}{\vc{C}_{\perp}^2+\nu^2}\mcdot\left(\frac{\vc{B}_{\perp}}{\vc{B}_{\perp}^2+\nu^2}-\frac{\vc{A}_{\perp}}{\vc{A}_{\perp}^2+\nu^2}\right)\nonumber\\
 &&\hspace*{-1cm}\qquad \times \big(1-\cos[(\Omega_1-\Omega_3)\Delta z]\big)-2\,\frac{\vc{C}_{\perp}}{\vc{C}_{\perp}^2+\nu^2}\mcdot \frac{\vc{B}_{\perp}}{\vc{B}_{\perp}^2+\nu^2}  \mcdot \big(1-\cos[(\Omega_2-\Omega_3)\Delta z]\big) \nn\\
 &&\hspace*{-1cm}\qquad +2\,\frac{\vc{A}_{\perp}}{\vc{A}_{\perp}^2+\nu^2}\mcdot\left(\frac{\vc{A}_{\perp}}{\vc{A}_{\perp}^2+\nu^2}-\frac{\vc{D}_{\perp}}{\vc{D}_{\perp}^2+\nu^2}\right)\big(1-\cos[\Omega_4\Delta z]\big) \nn \\
 &&\hspace*{-1cm}\qquad +2\,\frac{\vc{A}_{\perp}}{\vc{A}_{\perp}^2+\nu^2}\mcdot \frac{\vc{D}_{\perp}}{\vc{D}_{\perp}^2+\nu^2}\big(1-\cos[\Omega_5\Delta z]\big)\Bigg) \Bigg] \nn \\
 &&\hspace*{-1cm}\qquad +\, m^2 \Bigg[  2\, \frac{1}{\vc{B}_{\perp}^2+\nu^2} \mcdot \left(\frac{1}{\vc{B}_{\perp}^2+\nu^2}-\frac{1}{\vc{A}_{\perp}^2+\nu^2}\right) \big( 1-\cos[(\Omega_1 -\Omega_2)\Delta z]  \big) +\, \ldots \Bigg]\Bigg\} \, .
\label{eq:fullmassive2} 
\end{eqnarray}
Here, the ellipses denote again analogous terms as in the first square bracket following the pattern as indicated, and $\nu=m$ as given in Eq.~\eqref{eq:nu3}. Note that in all three cases, the massive vacuum splitting functions given in Eqs. (\ref{eq:massiveQQg}) and (\ref{eq:massivegQQbar}) do not factor out as it was the case for the massless in-medium splitting functions.

\subsection{Soft gluon approximation}

In this section, we consider the soft gluon approximation (SGA) of the full massive splitting functions in Eqs.~(\ref{eq:fullmassive1}) and (\ref{eq:fullmassive2}) by taking the limit $x\to 0$. We then make the connection with the results obtained in the traditional picture of parton energy loss~\cite{Djordjevic:2003zk}. For comparison, we first present the massless results in the SGA as derived in~\cite{Ovanesyan:2011xy,Ovanesyan:2011kn} and earlier in~\cite{Gyulassy:2000er} using the traditional approach of parton energy loss:
\begin{eqnarray}
 && \!\!\!\!\!\!\!  x \left( \frac{dN^{\mathrm{SGA}}}{ dxd^2{\vc k}_\perp}\right)_{ \left\{ \begin{array}{c}   q \rightarrow qg \\    
 g\rightarrow gg  \\  g \rightarrow q\bar{q} \\ q \rightarrow gq\\   \end{array} \right\} }  =  \frac{\alpha_s}{\pi^2} 
  \left\{ \begin{array}{c}   C_F[ 1+ {\cal O}(x) ] \\  C_A[ 1+ {\cal O}(x) ]  \\
  T_R [0 + \frac{x}{2} + {\cal O}(x^2)]  \\    C_F [0 + \frac{x}{2} + {\cal O}(x^2)]
  \end{array} \right\} \ \times    \int {d\Delta z}   \left\{ \begin{array}{c}    \frac{1}{\lambda_g(z)} \\[1ex]  \frac{1}{\lambda_g(z)}  \\
 \frac{1}{\lambda_q(z)}  \\  \frac{1}{\lambda_q(z)}    \end{array} \right\}
\int  d^2{\vc q}_\perp  \;  \nonumber \\
&&  \qquad \qquad  \qquad \qquad \qquad  \times\frac{1}{\sigma_{el}} \frac{d\sigma_{el}^{\; {\mathrm{med}}}}{d^2 {\vc q}_\perp}  \frac{2 \vc{k}_{\perp}  \cdot \vc{q}_{\perp} }{\vc{k}_{\perp}^2 (\vc{k}_{\perp}-\vc{q}_{\perp})^2}
   \left [ 1-\cos \frac{   (\vc{k}_{\perp}-\vc{q}_{\perp})^2}{xp^+_0} \Delta z \right]. \nn \\
\label{smallx}
\end{eqnarray} 
Note that we keep the ${\cal O}(x)$ expressions for the off-diagonal splitting functions even though they actually vanish for strictly $x\to 0$. 

Next, we consider the massive splitting functions in the SGA. We would like to point out that there is some ambiguity for defining the massive small-$x$ result. This ambiguity arises for the diagonal splitting process $Q\to Qg$, where the mass term is proportional to $x^2m^2$. This means that any mass dependence vanishes for strictly $x\to 0$. Therefore, in order to keep a finite mass correction even for $x\to 0$, one conventionally chooses to keep the first order correction of the mass in the denominator. However, this convention leaves some ambiguity at what stage of the derivation one should keep the first order correction in the denominator. When deriving the SGA, one ends up with the following expression
\be\label{eq:djordjevic}
\frac{\vc{k}_\perp-\vc{q}_\perp}{(\vc{k}_{\perp}-\vc{q}_{\perp})^2+x^2m^2}\left(\frac{\vc{k}_\perp-\vc{q}_\perp}{(\vc{k}_{\perp}-\vc{q}_{\perp})^2+x^2m^2}-\frac{\vc{k}_\perp}{\vc{k}_{\perp}^2+x^2m^2}\right) \, ,
\ee
where we kept the mass terms $\sim x^2m^2$ in the denominators. This structure is eventually multiplied by a factor involving a cosine similar to the massless case in Eq.~(\ref{smallx}). Keeping the masses at this stage would be consistent with the convention in~\cite{Djordjevic:2003zk}, where the massive small-$x$ result was derived within the conventional approach to parton energy loss. In~\cite{Djordjevic:2003zk}, the final result is cast in the following form
\be
\frac{\vc{k}_{\perp}  \cdot \vc{q}_{\perp} (\vc{k}_\perp - \vc{q}_\perp)^2+x^2m^2\vc{q}_\perp\cdot(\vc{q}_\perp-\vc{k}_\perp)}{[\vc{k}_{\perp}^2+x^2m^2][(\vc{k}_{\perp}-\vc{q}_{\perp})^2+x^2m^2]^2} \, .
\ee
Note that this result involves three factors in the denominator which is different than in the massless case. However, this expression can also be written as
\be\label{eq:sga3}
\frac{\vc{k}_\perp\cdot\vc{q}_\perp}{[\vc{k}_{\perp}^2+x^2m^2][(\vc{k}_{\perp}-\vc{q}_{\perp})^2+x^2m^2]} -\f{x^2m^2\, \vc{q}_\perp\cdot(2\vc{k}_\perp-\vc{q}_\perp)}{[\vc{k}_{\perp}^2+x^2m^2][(\vc{k}_{\perp}-\vc{q}_{\perp})^2+x^2m^2]^2}\, ,
\ee
where now the first term has a similar structure as the massless result in Eq.~(\ref{smallx}). The second term is proportional $\sim x^2m^2$ and vanishes for strictly $x\to 0$. Therefore, we choose to keep only the first term in Eq.~(\ref{eq:sga3}) in the SGA. This version of the massive SGA for $Q\to Qg$ is more similar to the massless result and more importantly, it is consistent with the results for the off-diagonal splitting functions in the SGA where there are no ambiguities. Note that for $g\to Q\bar Q$, the first mass correction has no $x$-dependence, see~Eq.~(\ref{eq:fullmassive2}) and also Eq.~(\ref{smallx3}) below. For $Q\to gQ$, the mass correction is proportional $\sim(1-x)^2m^2$ which also becomes $\sim m^2$ for $x\to 0$. Since there is no ambiguities for the off-diagonal small-$x$ results, we choose to define the massive diagonal SGA result for $Q\to Qg$ in analogy to them. The complete result for $Q\to Qg$ in the SGA is
\begin{eqnarray}
 && \!\!\!\!\!\!\!  x \left( \frac{dN^{\mathrm{SGA}}}{dxd^2{\vc k}_\perp}\right)_{Q \rightarrow Qg } =  \frac{\alpha_s}{\pi^2} C_F
  \int {d\Delta z}    \frac{1}{\lambda_g(z)} 
\int  d^2{\vc q}_\perp  \frac{1}{\sigma_{el}} \frac{d\sigma_{el}^{\; {\mathrm{med}}}}{d^2 {\vc q}_\perp} \;  \nonumber \\
&&  \qquad \qquad \times \frac{2 \vc{k}_{\perp}  \cdot \vc{q}_{\perp} }{[\vc{k}_{\perp}^2+x^2m^2][(\vc{k}_{\perp}-\vc{q}_{\perp})^2+x^2m^2]}
   \left [ 1-\cos \frac{   (\vc{k}_{\perp}-\vc{q}_{\perp})^2+x^2m^2}{xp^+_0} \Delta z \right]. \nn \\
\label{smallx1}
\end{eqnarray} 
Note that we also keep a finite mass correction in the phase of the cosine. We would like to stress again that this convention is different than the one chosen in~\cite{Djordjevic:2003zk} where the structure in Eq.~(\ref{eq:djordjevic}) was used instead. 

We now continue presenting the results for the off-diagonal splitting processes. In the small-$x$ limit, we have the following result for $Q\to gQ$
\begin{eqnarray}
 && \!\!\!\!\!\!\!  x \left( \frac{dN^{\mathrm{SGA}}}{dxd^2{\vc k}_\perp}\right)_{Q \rightarrow gQ } =  \frac{\alpha_s}{\pi^2} C_F\left(\f{x}{2}\right)
  \int {d\Delta z}    \frac{1}{\lambda_q(z)} 
\int d^2{\vc q}_\perp  \frac{1}{\sigma_{el}} \frac{d\sigma_{el}^{\; {\mathrm{med}}}}{d^2 {\vc q}_\perp} \;  \nonumber \\
&&  \qquad \qquad \times \frac{2 \vc{k}_{\perp}  \cdot \vc{q}_{\perp} }{[\vc{k}_{\perp}^2+m^2][(\vc{k}_{\perp}-\vc{q}_{\perp})^2+m^2]}
   \left [ 1-\cos \frac{   (\vc{k}_{\perp}-\vc{q}_{\perp})^2+m^2}{xp^+_0} \Delta z \right] \, , \nn \\
\label{smallx2}
\end{eqnarray} 
and for $g\to Q\bar Q$
\begin{eqnarray}
 && \!\!\!\!\!\!\!  x \left( \frac{dN^{\mathrm{SGA}}}{dxd^2{\vc k}_\perp}\right)_{g \rightarrow Q\bar Q } =  \frac{\alpha_s}{\pi^2} T_R\left(\f{x}{2}\right)
  \int {d\Delta z}    \frac{1}{\lambda_q(z)} 
\int d^2{\vc q}_\perp  \frac{1}{\sigma_{el}} \frac{d\sigma_{el}^{\; {\mathrm{med}}}}{d^2 {\vc q}_\perp} \;  \nonumber \\
&&  \qquad \qquad \times \frac{2 \vc{k}_{\perp}  \cdot \vc{q}_{\perp} }{[\vc{k}_{\perp}^2+m^2][(\vc{k}_{\perp}-\vc{q}_{\perp})^2+m^2]}
   \left [ 1-\cos \frac{   (\vc{k}_{\perp}-\vc{q}_{\perp})^2+m^2}{xp^+_0} \Delta z \right].
\label{smallx3}
\end{eqnarray} 
Note that for the two off-diagonal splitting functions, the mass correction is directly $\sim m^2$ without any dependence on $x$ as discussed above.

\subsection{Numerical results}

\begin{figure*}[t!]
\begin{center}
\vspace*{10mm}
\includegraphics[width=0.4\textwidth,trim=1cm 2cm 1cm 0.5cm ]{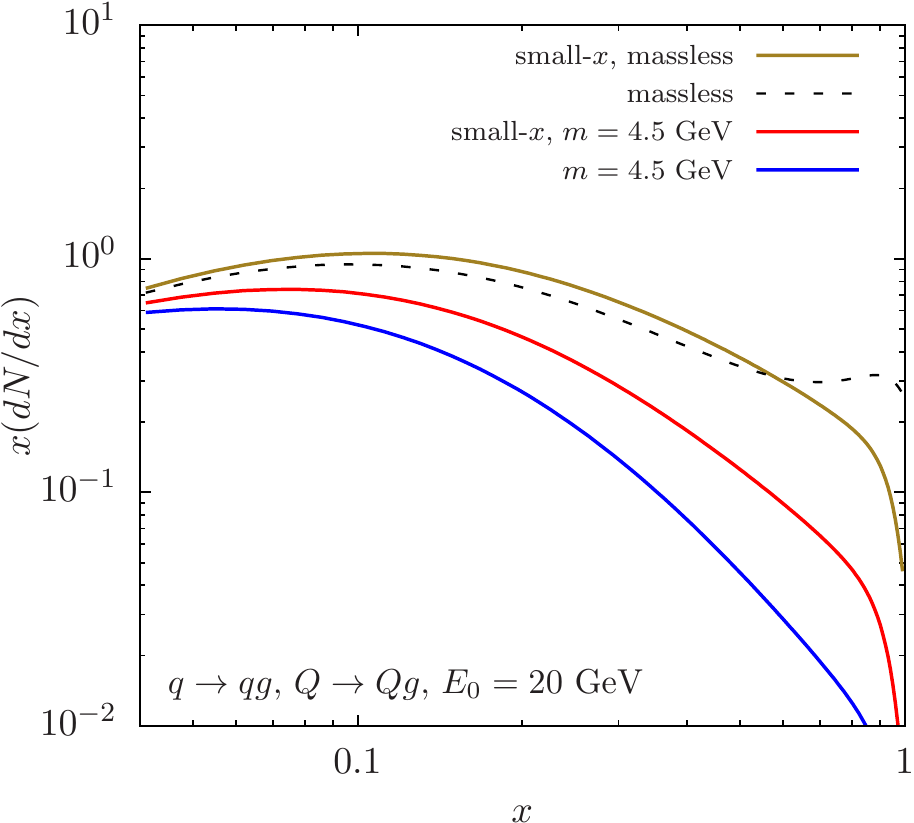} 
\hspace*{2cm}
\includegraphics[width=0.4\textwidth,trim=1cm 2cm 1cm 0.5cm ]{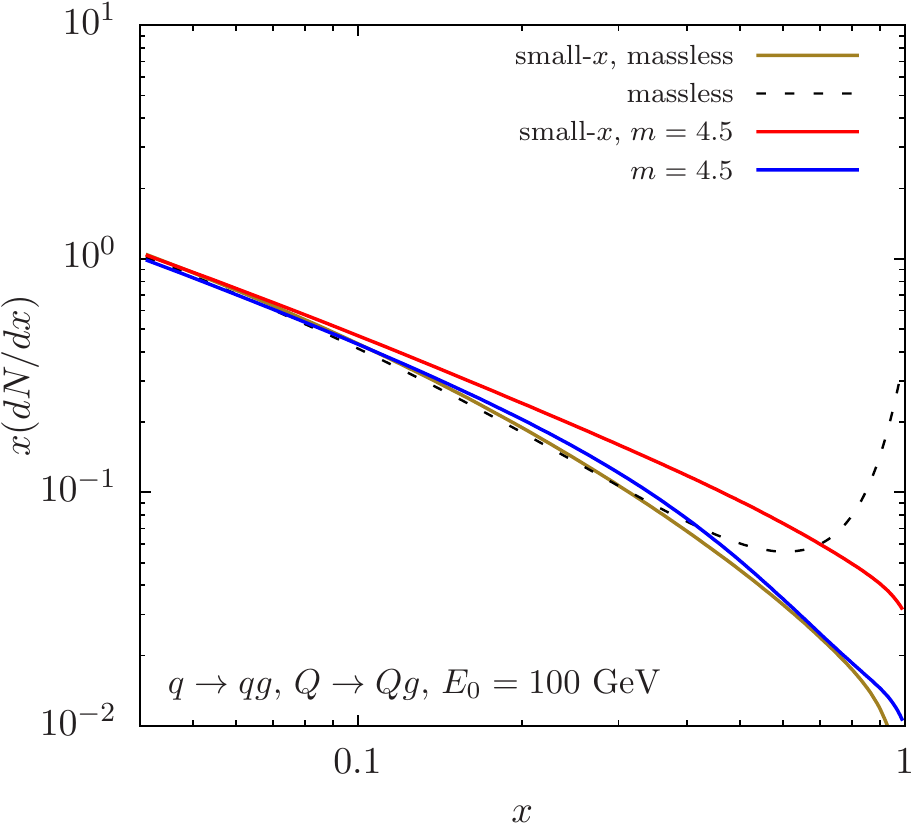} 
\end{center}
\vspace*{1.cm}
\caption{\label{fig:qq100} Comparison of the intensity spectra $x(dN/dx)$ for the quark-to-quark splitting process. The massive results for the full splitting function $Q\to Qg$ are shown in blue, whereas the corresponding small-$x$ results are shown in red. We choose the mass of the bottom quark as an example, $m_b=4.5$~GeV. For comparison, we also plot the massless results $q\to qg$ for both the full splitting function (dashed black) and the small-$x$ limit (green). The $\vc{q}_\perp$ and $\Delta z$ integrals are evaluated numerically with a realistic model for the medium and physical phase space cuts, see text and~\cite{Chien:2015vja}. As an example, we choose the incident parent parton energy as $E_0=p_0^+/2=20$~GeV (left) and $E_0=100$~GeV (right).}
\end{figure*}

\begin {figure*}[t]
\begin{center}
\vspace*{10mm}
\includegraphics[width=0.4\textwidth,trim=1cm 2cm 1cm 0.5cm ]{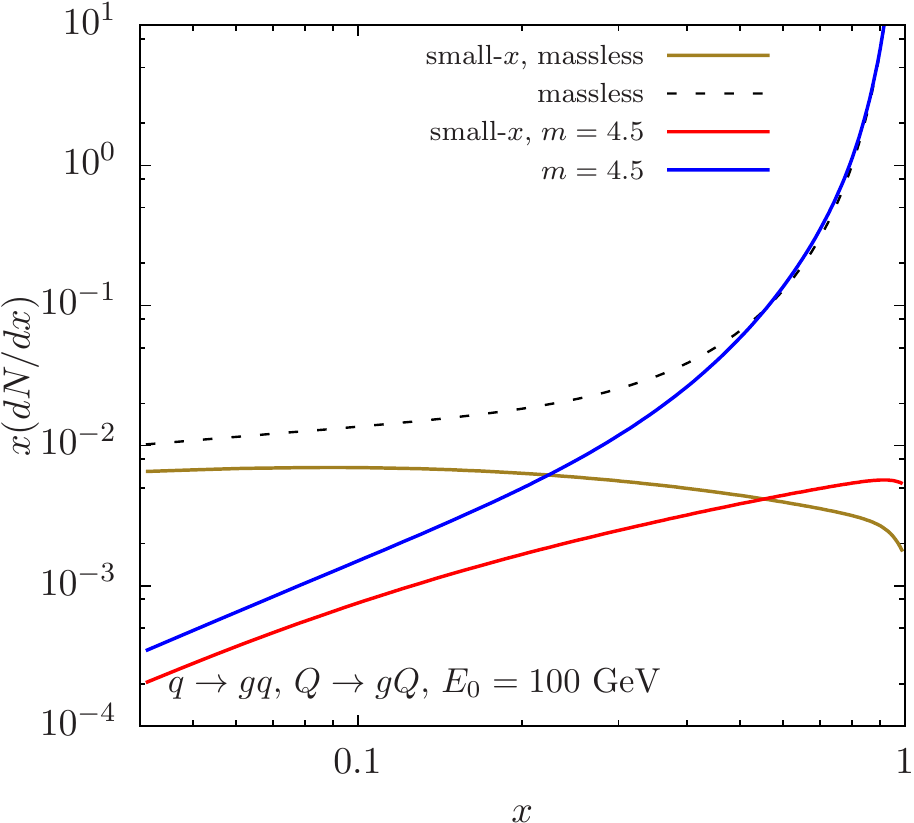} 
\hspace*{2cm}
\includegraphics[width=0.4\textwidth,trim=1cm 2cm 1cm 0.5cm ]{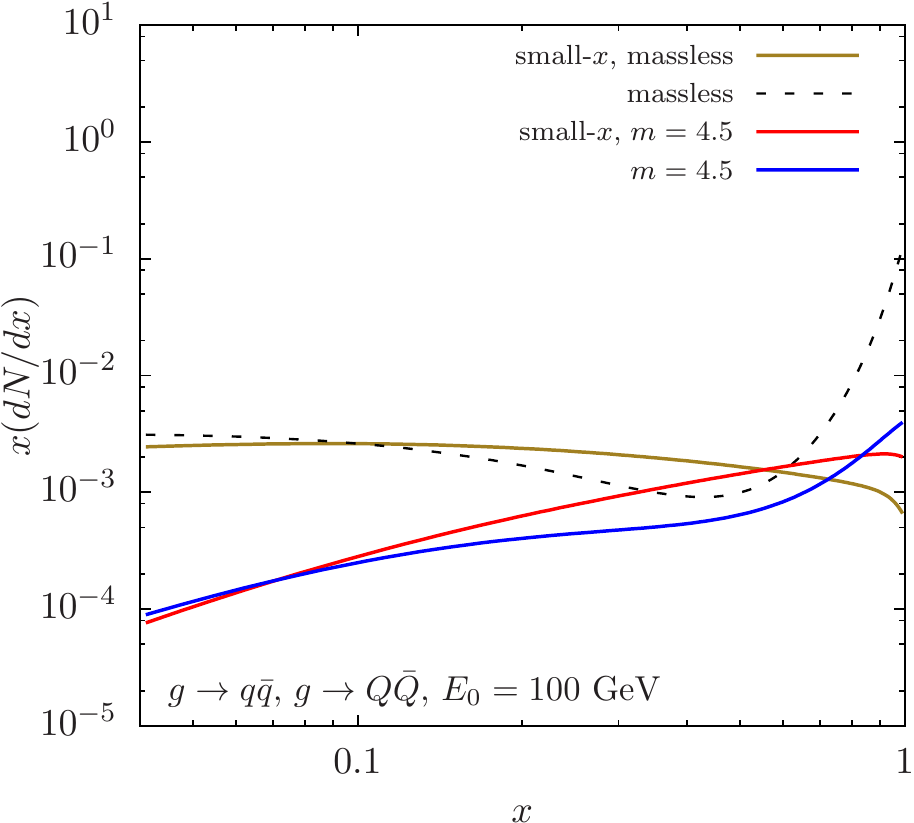} 
\end{center}
\vspace*{1.cm}
\caption{\label{fig:offdiagsplit} Similar to Fig.~\ref{fig:qq100} but for the off-diagonal splitting processes $Q\to gQ$ (left) and $g\to Q\bar Q$ (right) for $E_0=100$~GeV.}
\end{figure*}

In this section, we present numerical results for the massive in-medium splitting kernels. We compare the full splitting kernels with the soft gluon approximated results and study the finite mass effects. We perform the $\vc{q}_\perp$ and $\Delta z$ integrations in Eqs.~(\ref{eq:fullmassive1}), (\ref{eq:fullmassive2}) and~(\ref{smallx1})--(\ref{smallx3}) numerically using a realistic model for the medium. For details of the medium properties, see the appendix of \cite{Chien:2015vja}. Here, as an example, we present the results for the QGP produced in central Pb+Pb collisions at $\sqrt{s_\mathrm{NN}}=5.02$ TeV at the LHC. We study the case for an incident bottom quark with mass $m_b=4.5$ GeV, and choose the coupling between the hard partons and the QGP medium $g=2.0$. The overall mass effects are smaller for a charm quark mass of $m_c=1.3$~GeV, nevertheless we find qualitatively similar results. 

In Figs.~\ref{fig:qq100} and~\ref{fig:offdiagsplit}, we show the results for the intensity spectra $x(dN/dx)$, which are obtained by integrating over $\vc{k}_\perp$ up to $\vc{k}_{\perp,\mathrm{max}}=2E_0\sqrt{x(1-x)}$. This choice is just for illustrational purposes. In section~\ref{sec:three} below, we consider a different upper integration limit for $\vc{k}_\perp$ that is required by how the in-medium splitting functions have to be treated when they appear in an actual cross section. In Fig.~\ref{fig:qq100}, we consider two initial parton energies $E_0=p_0^+/2=20$~GeV (left) and $E_0=100$~GeV (right). We show the full massive splitting function for $Q\to Qg$ in blue, whereas the corresponding soft gluon approximated result is shown in red. For comparison, we also plot the massless results $q\to qg$ for both the full splitting function (dashed black) and the small-$x$ limit (green). Note that we take into account a finite Debye mass $m_D = gT$ for the medium, which appear for both the massless and the massive cases. It can be seen clearly that the mass effects are a large-$x$ effect since all four curves for both choices of $E_0$ in Fig.~\ref{fig:qq100} are very close together at small-$x$. This is to be expected as the mass corrections are of the form $\sim x^2m^2$ for $Q\to Qg$. By comparing the two results for the full splitting functions for massive quarks (blue) and massless quarks (dashed black), one finds a significant difference in the large-$x$ region for $x>0.4$. Interestingly, the rise of the massless result at large-$x$ completely disappears when considering a finite bottom quark mass. As expected, the finite mass results are more relevant for $E_0=20$~GeV (left), where the differences are clearly larger.

In Fig.~\ref{fig:offdiagsplit}, we present analogous numerical results for the off-diagonal splitting functions $Q\to gQ$ (left) and $g\to Q\bar Q$ (right) for $E_0=100$~GeV. The finite mass effects are even more pronounced here than for the diagonal splitting $Q\to Qg$, and can be relevant for both the large and the small-$x$ region. The enhanced effect in the small-$x$ region is consistent with the fact that the mass corrections for the processes $Q\to gQ$ and $g\to Q\bar Q$ are proportional $\sim (1-x)^2m^2$ and $\sim m^2$ respectively, and thus remain finite when taking $x\to 0$. Although the mass corrections can be large in the small-$x$ region, it is instructive to keep in mind that both off-diagonal splitting functions vanish when $x\to 0$, as can be seen clearly from Eqs.~\eqref{smallx2} and \eqref{smallx3}. Therefore, the overall numerical impact of the finite mass effects at the level of $x(dN/dx)$ from these regions is not directly translated to the cross section in heavy-ion collisions.

In summary, we find that finite mass effects are indeed very significant at the level of the intensity spectra $x(dN/dx)$. Eventually this can have a sizable numerical impact for the suppression of heavy mesons in heavy ion collisions as discussed in the next section.

\section{Application to $\mathrm{PbPb}\to HX$ at NLO \label{sec:three}}

In this section, we first introduce a new framework for including in-medium effects consistent with next-to-leading order calculations in QCD for inclusive hadron production in heavy ion collisions. This can be achieved by making use of the in-medium massive slitting functions derived in last section and by effectively introducing in-medium fragmentation functions. We then consider the cross section for open heavy flavor production in proton-proton collisions, and provide numerical results in the so-called zero mass variable flavor number scheme (ZM-VFNS). Finally, we present results for the suppression of heavy meson production in Pb+Pb collisions for both $\sqrt{s_\mathrm{NN}}=5.02$~TeV and 2.76~TeV and compare to the experimental data at the LHC.

\subsection{In-medium fragmentation functions}

\begin {figure*}[t]
\begin{center}
\vspace*{10mm}
\includegraphics[width=\textwidth, trim = 0cm 2.5cm 0cm 2.5cm]{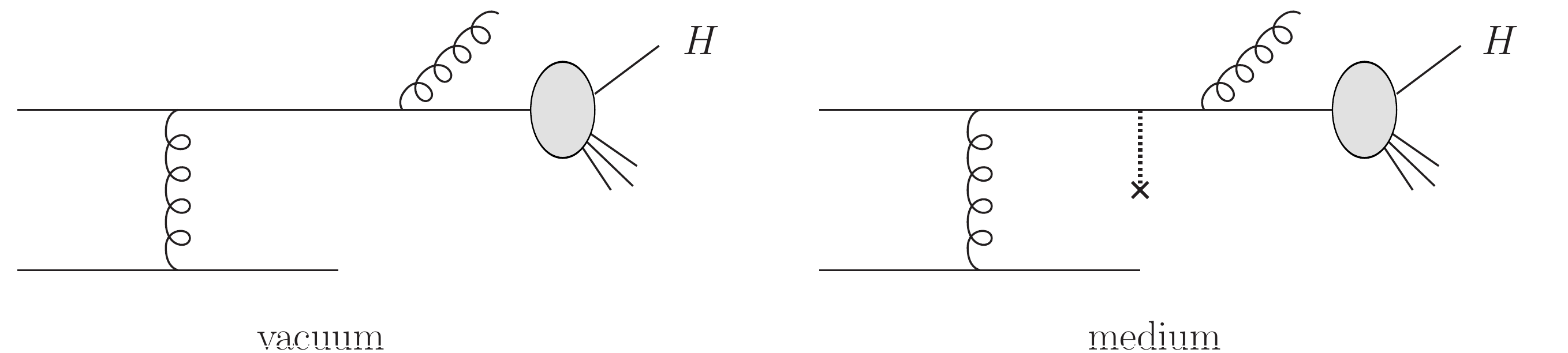} 
\end{center}
\vspace*{1.cm}
\caption{\label{fig:vacmedNLO} Real-emission corrections at next-to-leading order in QCD for inclusive hadron production. The vacuum case (proton-proton) is shown on the left hand side and the medium induced diagram (Pb+Pb) is shown on the right. The gray ellipses represent the standard vacuum fragmentation functions for the inclusive production of a hadron $H$. The dotted line represents the interaction with the medium. See text for further discussions.}
\end{figure*}

In this section we derive a framework to include in-medium interactions for $\mathrm{PbPb}\to HX$ which is consistent with NLO calculations in the vacuum for $pp\to HX$. Initial state Cold Nuclear Matter (CNM) effects will be included only for numerical evaluations at the very end, with the actual implementation explained in more details in \cite{Chien:2015vja}. An improved treatment of CNM energy loss using SCET$_\mathrm{G}$-based initial-state splitting functions~\cite{Ovanesyan:2015dop} will be left for future work. The framework that we develop in this section is related to jet calculations in~\cite{Chien:2015hda,jet_frag} and to some extend it corresponds to a first order expansion of the DGLAP formalism developed in~\cite{Kang:2014xsa,Chien:2015vja}. For a discussion of a medium-modified DGLAP in semi-inclusive DIS, see~\cite{Chang:2014fba}.

Interactions with the hot and dense QCD medium affect partons after the hard-scattering event but before they eventually fragment into hadrons. To NLO in the strong coupling constant, we have to consider one-loop real and virtual corrections for the outgoing final state parton. As an example, we consider the corrections to the leading-order hard process $qq\to qq$ as shown in Fig.~\ref{fig:vacmedNLO}. 
In the vacuum, a splitting process such as that shown in Fig.~\ref{fig:vacmedNLO} (left) needs to be taken into account. Such a contribution will eventually lead to the DGLAP evolution of the vacuum fragmentation function. On the other hand, in the QCD medium, besides the vacuum splitting process, an medium-induced splitting process as shown in Fig.~\ref{fig:vacmedNLO} (right) will also happen, which leads to additional contributions to the cross section for the hadron production in heavy ion collisions.   
Of course, when squaring the amplitude corresponding to the medium-induced diagram on the right hand side of Fig.~\ref{fig:vacmedNLO}, we actually need take into account all relevant single- and double-Born diagrams to first order in opacity as discussed in section~\ref{sec:two}. The gray ellipses denote the standard vacuum fragmentation function for both situations. We start by rederiving the vacuum case and we then continue by describing how this calculation can be extended to the medium case. At one-loop order, the relevant part that describes the splitting of the final state parton can be schematically written as
\be\label{eq:sigmedium}
\sum_j\hat \sigma^{(0)}_{i} \otimes P_{ji} \otimes D_j^{H}\, .
\ee
Here, $\hat \sigma^{(0)}_i$ is the leading-order hard-scattering cross section to produce a parton $i$, $P_{ji}$ is the leading-order Altarelli-Parisi splitting function for $i\to j$ and $D_j^H$ is the parton-to-hadron fragmentation function. This generic structure in Eq.~(\ref{eq:sigmedium}) can be found in standard textbooks such as~\cite{Halzen:1984mc}. The symbols $\otimes$ denote convolution products. This structure can be obtained by calculating a parton-to-parton fragmentation function to one-loop order or by considering the relevant splitting process to be part of the hard-scattering function, see~\cite{Aversa:1988vb,Jager:2002xm}. Note that these two possibilities are fully equivalent to first order. We choose to present the calculation for a parton-to-parton fragmentation function. Conceptually, we want to treat the two splitting processes shown in Fig.~\ref{fig:vacmedNLO} (vacuum and medium case) to be part of the first order correction to the leading-order process $qq\to qq$. 

We start by calculating the massless partonic quark and gluon fragmentation functions in the vacuum. 
Massive in-medium splitting functions can be implemented in a straightforward way as well. We will comment on the extension to massive quarks below.
For $q\to q$ and $g\to g$, we need to take into account both real and virtual corrections. Using the method of~\cite{Collins:1988wj}, one can express the contributions of the virtual graphs in terms of splitting functions derived from real emission graphs. This is consistent with the so-called flavor and momentum sum rules~\cite{Altarelli:1977zs}. From here on, we switch to the more traditional convention, where for any given splitting process, the radiated parton carries a momentum fraction $1-z$ instead of $z$ as in the previous section. For $q\to q$, we have
\ba\label{eq:vacq}
D_q^{q,(1),\mathrm{vac}}(z,\mu) & = & \f{\as}{\pi}C_F\int_{Q_0}^{\mu}\f{dk_\perp}{k_\perp}\f{1+z^2}{1-z}-\f{\as}{\pi}C_F\delta(1-z)\int_{Q_0}^{\mu}\f{dk_\perp}{k_\perp}\int_0^1 dx \f{1+x^2}{1-x} \nn \\
& = & \f{\as}{\pi}C_F\int_{Q_0}^{\mu}\f{dk_\perp}{k_\perp}\left(\f{1+z^2}{1-z}\right)_+ \, ,
\ea
where we use the notation $k_\perp=|\vc{k}_\perp|$, and $k_\perp$ is integrated between a lower scale $Q_0$ and an upper cutoff $\mu$ that is usually identified as the relevant hard scale of the process in consideration. If one takes the derivative of Eq.~(\ref{eq:vacq}) with respect to the upper integration limit $\mu$, one will derive the DGLAP evolution equations. Note that the integral over $x$ in the first line is divergent by itself but it is cancelled between real and virtual contributions, and we are left with a regularized plus distribution in the second line. 

For the two off-diagonal fragmentation functions at one-loop order, we have
\ba
D_g^{q,(1),\mathrm{vac}}(z,\mu) & = &  \f{\as}{\pi}C_F\int_{Q_0}^{\mu}\f{dk_\perp}{k_\perp}\f{1+(1-z)^2}{z} \, , \\
D_q^{g,(1),\mathrm{vac}}(z,\mu) & = &  \f{\as}{\pi}T_F\int_{Q_0}^{\mu}\f{dk_\perp}{k_\perp}(z^2+(1-z)^2) \, .
\ea
The process $g\to g$ is slightly more involved and also needs special attention in the medium case
\ba\label{eq:vacg}
D_g^{g,(1),\mathrm{vac}}(z,\mu) & = & \f{\as}{\pi}2C_A\int_{Q_0}^{\mu}\f{dk_\perp}{k_\perp}\left(\f{z}{1-z}+\f{1-z}{z}+z(1-z)\right)\nn \\
&& -\f{\as}{\pi}\f{\delta(1-z)}{2} \int_{Q_0}^{\mu}\f{dk_\perp}{k_\perp}\int_0^1 dx \left[ 2C_A \left(\f{x}{1-x}+\f{1-x}{x}+x(1-x)\right) \right. \nn \\
&& +2N_fT_F(x^2+(1-x)^2)\Big] \nn \\
& = & \f{\as}{\pi}2C_A\int_{Q_0}^{\mu}\f{dk_\perp}{k_\perp}\left\{\left(\f{z}{(1-z)}_+ +\f{1-z}{z}+z(1-z)\right) +\f{\beta_0}{2}\delta(1-z) \right\} \, , \nn \\
\ea
where $\beta_0=11/3 C_A-4/3 T_F N_f$. Note that the expressions in the second and third lines correspond to virtual corrections for both gluon and quark loops. The last line is obtained by utilizing the definition of the plus function. Again all the divergences cancel between real and virtual corrections, and we are left with a regularized plus distribution. We have now obtained the standard expressions, where the partonic fragmentation functions are written in terms of the leading-order Altarelli-Parisi splitting functions. 

For notational convenience and easy generalization to the medium case, let us introduce the functions ${\cal P}_{i\to jk}(z,\mu)$ for every splitting process $i\to jk$, where $k$ corresponds to the emitted parton carrying away the momentum fraction $1-z$. The functions ${\cal P}_{i\to jk}(z,\mu)$ are related to the splitting functions $(dN/dz/d^2k_\perp)_{i\to jk}$ defined in section~\ref{sec:two} as
\be\label{eq:defP}
{\cal P}_{i\to jk}(z,\mu) = \int_{Q_0}^\mu dk_\perp \left(\f{dN}{dzdk_\perp}\right)_{i\to jk} = \int_{Q_0}^\mu dk_\perp 2\pi k_\perp \left( \f{dN}{dzd^2k_\perp}\right)_{i\to jk}  \, .
\ee
We use this identification both for the vacuum and the medium case. To be specific, we always include a superscript ``vac'' or ``med'' below. Note that we included the $k_\perp$ integral in the definition of ${\cal P}_{i\to jk}(z,\mu)$ as it is always the same. For example, for the splitting process $q\to qg$ in the vacuum, we have
\be
{\cal P}^{\mathrm{vac}}_{q\to qg}(z,\mu)= \f{\as}{\pi}C_F \int_{Q_0}^\mu \f{dk_\perp}{k_\perp}\f{1+z^2}{1-z} \, .
\ee
Using this notation, we can now write the partonic vacuum fragmentation functions derived above as
\begin{subequations}
\label{eq:Dvac}
\bea
D_q^{q,(1),\mathrm{vac}}(z,\mu) = &  {\cal P}^{\mathrm{vac}}_{q\to qg}(z,\mu)-\delta(1-z)\int_0^1 dx\,  {\cal P}^{\mathrm{vac}}_{q\to qg}(x,\mu)\,, \\
D_g^{g,(1),\mathrm{vac}}(z,\mu) = & {\cal P}^{\mathrm{vac}}_{g\to gg}(z,\mu)-\f{\delta(1-z)}{2} \int_0^1 dx \left[ {\cal P}^{\mathrm{vac}}_{g\to gg}(x,\mu)+2N_f{\cal P}^{\mathrm{vac}}_{g\to q\bar q}(x,\mu)\right] \, , \\
D_g^{q,(1),\mathrm{vac}}(z,\mu)  = & {\cal P}^{\mathrm{vac}}_{q\to gq}(z,\mu) \, , \\
D_q^{g,(1),\mathrm{vac}}(z,\mu)  = & {\cal P}^{\mathrm{vac}}_{g\to q\bar q}(z,\mu) \, ,
\eea
\end{subequations}
which contain both real and virtual contributions.

With such notations, it is straightforward to extend these results to the medium case, where one has to consider both the vacuum splitting function as well as a medium induced part. Therefore, we can directly make the following substitutions in Eq.~(\ref{eq:Dvac}) above
\be
{\cal P}^{\mathrm{vac}}_{i\to jk}(z,\mu)\to {\cal P}_{i\to jk}(z,\mu)={\cal P}^{\mathrm{vac}}_{i\to jk}(z,\mu)+{\cal P}^{\mathrm{med}}_{i\to jk}(z,\mu) \, .
\ee
Here, the ${\cal P}^{\mathrm{vac}}_{i\to jk}(z,\mu)$ are the vacuum splitting functions as discussed above and ${\cal P}^{\mathrm{med}}_{i\to jk}(z,\mu)$ are the in-medium splitting functions as derived in section~\ref{sec:two} and integrated over $k_\perp$ as defined in Eq.~(\ref{eq:defP}). Eventually, we need to ``match'' onto the standard vacuum fragmentation functions. We now continue by evaluating this matching procedure and calculate the convolution with the standard fragmentation functions. Schematically, we have the following structure for the medium case
\be\label{eq:medNLO}
\sum_j\hat \sigma^{(0)}_{i} \otimes {\cal P}^{\mathrm{med}}_{i\to jk} \otimes D_j^{H} \equiv \hat \sigma^{(0)}_i \otimes D_i^{H,\mathrm{med}}\, .
\ee
Effectively, the functions $D_i^{H,\mathrm{med}}(z,\mu)$ can be considered as medium-modified fragmentation functions~\cite{Vitev:2002pf,Guo:2000nz}. Even though we consider the medium induced splittings as a correction to the vacuum hard-scattering function, it is notationally more convenient to think of it as a medium-modified fragmentation function. The medium-modified FF will then be convolved with the leading-order hard-scattering cross section. As it is formally a one-loop correction, we are then going to add it to the NLO calculation in the vacuum. 

Following the definition of the medium-modified quark and gluon fragmentation functions $D_i^{H,\mathrm{med}}$ in Eq.~(\ref{eq:medNLO}), we find
\begin{subequations}
\label{eq:Dmed}
\bea
D_q^{H,\mathrm{med}}(z,\mu)  = & \int_z^1\f{dz'}{z'}D_q^H\left(\f{z}{z'},\mu \right){\cal P}^{\mathrm{med}}_{q\to qg}(z',\mu)-D_q^H(z,\mu)\int_0^1dz' {\cal P}^{\mathrm{med}}_{q\to qg}(z',\mu) \nn \\
& +\int_z^1\f{dz'}{z'} D_g^H\left(\f{z}{z'},\mu \right) {\cal P}^{\mathrm{med}}_{q\to gq}(z',\mu) \, , 
\\
D_g^{H,\mathrm{med}}(z,\mu)  = &  \int_z^1\f{dz'}{z'}D_g^H\left(\f{z}{z'},\mu \right){\cal P}^{\mathrm{med}}_{g\to gg}(z',\mu)-\f{D_g^H(z,\mu)}{2}\int_0^1dz' \left[  {\cal P}^{\mathrm{med}}_{g\to gg}(z',\mu)\right.  \nn \\
&\left. +2 N_f {\cal P}^{\mathrm{med}}_{g\to q\bar q}(z',\mu)\right]  +\int_z^1\f{dz'}{z'} \sum_{i=q,\bar q} D_i^H\left(\f{z}{z'},\mu \right) {\cal P}^{\mathrm{med}}_{g\to q\bar q}(z',\mu) \, .
\eea
\end{subequations}
At this point one can make a direct connection between the new treatment of the medium effects as proposed here and the approach considered in~\cite{Kang:2014xsa,Chien:2015vja}. In these two papers, the authors derived medium-modified DGLAP equations. The DGLAP equations including medium effects can be obtained by taking the derivative of the above Eq.~\eqref{eq:Dmed} with respect to the scale $\mu$. Using medium-modified DGLAP equations essentially leads to an exponentiation of the in-medium branchings. In~\cite{Kang:2014xsa,Chien:2015vja}, a close connection was established between the medium-modified DGLAP equations and traditional parton energy loss calculations. In our case, we only consider the first order correction in $\alpha_s$ evaluated in the opacity series expansion. The numerical results of the two approaches turn out to be very similar but the approach proposed in this paper is easier to implement. In addition, the new approach has a close connection to NLO calculations in the vacuum which we use as proton-proton baseline as discussed in the next section. Differences between the two approaches are expected only when the observed hadrons have a relatively small transverse momentum $p_T$.

Although schematically correct and overall finite, Eq.~\eqref{eq:Dmed} cannot be used as they are since the individual terms can become numerically divergent at the phase space boundaries $z\to 0,~1$ due to the behavior of the splitting functions ${\cal P}_{i\to jk}(z,\mu)$, see Eqs.~\eqref{CohRadSX1}, \eqref{CohRadSX2}, and \eqref{eq:fullmassive1}. To rewrite the expressions into separate ``numerically stable'' pieces, we introduce plus distributions similar to \cite{Chien:2015vja}. For example, the quark in-medium fragmentation function can be directly written as
\bea\label{eq:finalq}
D_q^{H,\mathrm{med}}(z,\mu)  = & \int_z^1 \f{dz'}{z'} D_q^H\left(\f{z}{z'},\mu\right)\left[{\cal P}^{\mathrm{med}}_{q\to qg}(z',\mu) \right]_+ + \int_z^1  \f{dz'}{z'}  D_g^H\left(\f{z}{z'},\mu\right) {\cal P}^{\mathrm{med}}_{q\to gq}(z',\mu) \nn \\
 = & \int_z^1 dz' \left[ \f{1}{z'} D_q^H\left(\f{z}{z'},\mu\right) - D_q^H(z,\mu) \right]  {\cal P}^{\mathrm{med}}_{q\to qg}(z',\mu)  \nn \\
& - D_q^H(z,\mu)\int_0^z dz' {\cal P}^{\mathrm{med}}_{q\to qg}(z',\mu) + \int_z^1 \f{dz'}{z'}  D_g^H\left(\f{z}{z'},\mu\right) {\cal P}^{\mathrm{med}}_{q\to gq}(z',\mu) \, ,
\eea
which can be evaluated numerically. Note that this procedure can be applied to both the massless and the massive case. For $g\to g$, we have to separate off the vacuum splitting function from the in-medium result. We define
\be
{\cal P}^{\mathrm{med}}_{g\to gg}(z,\mu)=\left(\f{z}{1-z}+\f{1-z}{z}+z(1-z)\right) h^{\mathrm{med}}_{g\to gg}(z,\mu) \, .
\ee
We can now write the sum of both real- and virtual gluon contributions to the $g\to g$ splitting function as
\bea
&{\cal P}^{\mathrm{med}}_{g\to gg}(z,\mu) -\f{\delta(1-z)}{2} \int_0^1 dx\, {\cal P}^{\mathrm{med}}_{g\to gg}(x,\mu) = z\left[\f{h_{g\to gg}^{\mathrm{med}}(z,\mu)}{1-z} \right]_+ \nn \\
& +\left(\f{1-z}{z}+z(1-z)\right)h_{g\to gg}^{\mathrm{med}}(z,\mu)-\f{\delta(1-z)}{2}\int_0^1 dx (x(1-x)-2) \, h_{g\to gg}^{\mathrm{med}}(x,\mu),
\eea
where we used $h^{\mathrm{med}}_{g\to gg}(1-x,\mu)=h^{\mathrm{med}}_{g\to gg}(x,\mu)$. We choose a slightly different convention for expressing the plus distribution as in~\cite{Kang:2014xsa,Chien:2015vja}. All conventions are equivalent as long as the divergence in the $x$ integral is cancelled. The version here is the minimally required one, where only $h^{\mathrm{med}}_{g\to gg}(z,\mu)/(1-z)$ are written in terms of a plus distribution. Including the off-diagonal contribution, we can now write the in-medium fragmentation function $D_g^{H,\mathrm{med}}$ as
\bea
\label{eq:finalg}
D_g^{H,\mathrm{med}}(z,\mu)  = & \int_z^1dz'\left\{ \left[ D_g^H\left(\f{z}{z'},\mu\right)-D_g^H(z,\mu)\right]\f{h^{\mathrm{med}}_{g\to gg}(z',\mu)}{1-z'} \right.
\nn\\
&\hspace*{-1cm} \left.+\f{1}{z'}D_g^H\left(\f{z}{z'},\mu \right) \left[\f{1-z'}{z'}+z'(1-z')\right] h^{\mathrm{med}}_{g\to gg}(z',\mu) \right\}  
\nn \\
&\hspace*{-1cm}  - D_g(z,\mu) \left[\int_0^z dz' \f{h^{\mathrm{med}}_{g\to gg}(z',\mu)}{1-z'}+\f{1}{2} \int_0^1 dz' (z'(1-z')-2)h^{\mathrm{med}}_{g\to gg}(z',\mu) \right] \nn\\
&\hspace*{-1cm} -\f{D_g^H(z,\mu)}{2} 2N_f \int_0^1 dz' \, {\cal P}^{\mathrm{med}}_{g\to q\bar q}(z',\mu)+\int_z^1\f{dz'}{z'}\sum_{i=q,\bar q} D_i^H\left(\f{z}{z'},\mu \right) {\cal P}^{\mathrm{med}}_{g\to q\bar q}(z',\mu),
\eea
which can be evaluated numerically like the quark in-medium fragmentation function in Eq.~(\ref{eq:finalq}). Both quark and gluon in-medium FFs eventually need to be convolved with the leading-order quark and gluon production cross sections as shown in Eq.~(\ref{eq:medNLO}). The cross section in Pb+Pb collisions is obtained by adding the one-loop medium correction to the vacuum NLO result. In other words, we have
\bea
\label{eq:AA}
d\sigma^H_{\mathrm{PbPb}} = d\sigma^{H, {\rm NLO}}_{pp} + d\sigma^{H, {\rm med}}_{\mathrm{PbPb}},
\eea
where $d\sigma^{H, {\rm NLO}}_{pp}$ is the NLO cross section in the vacuum, and $d\sigma^{H, {\rm med}}_{\mathrm{PbPb}}$ is the one-loop medium correction, and schematically we have 
\bea
d\sigma^{H, {\rm med}}_{\mathrm{PbPb}} = \hat \sigma^{(0)}_i \otimes D_i^{H,\mathrm{med}}.
\eea

\subsection{Numerical results for $pp\to HX$ at NLO within the ZM-VFNS}
In this section we present numerical calculations for open heavy meson production in proton-proton collisions, $pp\to HX$. 
We choose to work within the ZM-VFNS~\cite{Collins:1986mp,Stavreva:2009vi}, in which one neglects all the heavy quark mass corrections in the partonic hard-scattering functions~\cite{Albino:2008fy,Accardi:2014qda}. Thus, this approximation is justified for $\mu\gg m_{c,b}$, where $\mu$ is the characteristic scale of the process and $m_{c,b}$ are the charm and bottom quark masses respectively. In the vacuum, the only characteristic scale of the process is the large transverse momentum of the produced hadron $\mu=p_T$. Therefore, the ZM-VFNS is expected to be applicable in the high-$p_T$ range: $p_T\gg m_{c,b}$. The exact range of validity of the ZM-VFNS needs to be checked by comparing theory and experimental data. For this reason, we perform several exemplary numerical calculations for $pp\to HX$ below.

On the other hand, the medium contribution is also sensitive to the properties of the QCD medium which introduces much lower energy scales than $p_T$. The characteristic scale for in-medium interactions is given by the typical momentum exchange between the incident parton and the QCD medium, which can be even smaller than the heavy quark mass. Therefore, while one can set the heavy quark masses to zero in the hard-scattering functions because of $p_T\gg m_{c,b}$, we do take into account the masses of both charm and bottom quarks in the medium-modified FFs for studying the medium contribution as discussed in the previous section~\ref{sec:two}. Such a set-up for the medium contribution is similar to \cite{Stavreva:2012aa}. 

\begin {figure*}[t]
\begin{center}
\vspace*{10mm}
\includegraphics[width=0.4\textwidth,trim=1cm 2cm 1cm 1cm ]{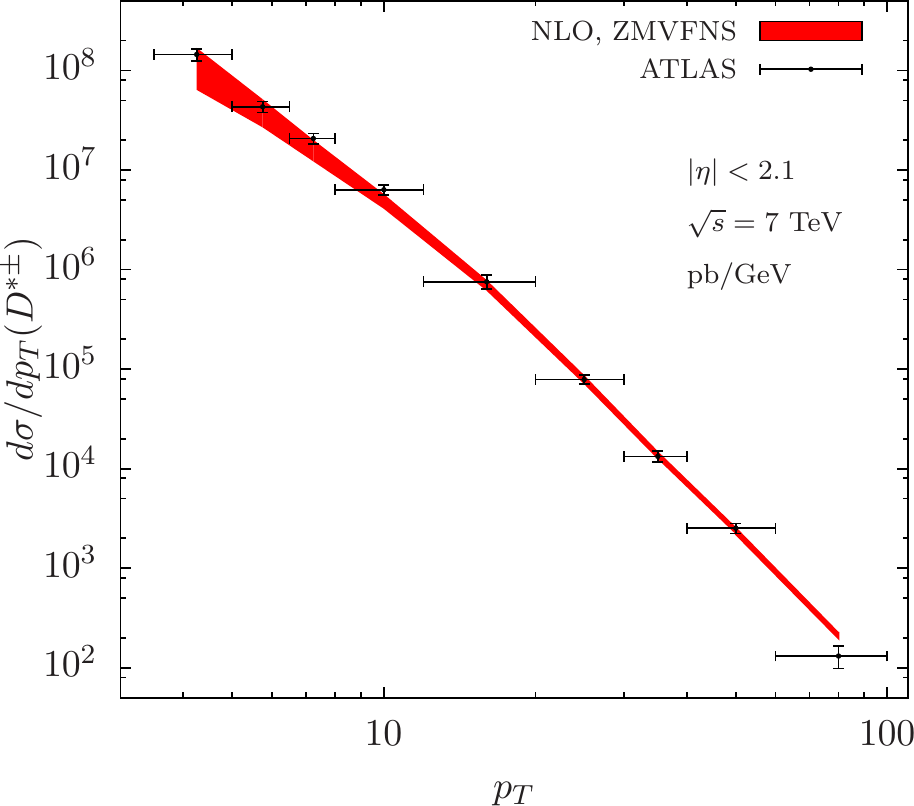} 
\hspace*{2cm}
\includegraphics[width=0.4\textwidth,trim=1cm 2cm 1cm 1cm ]{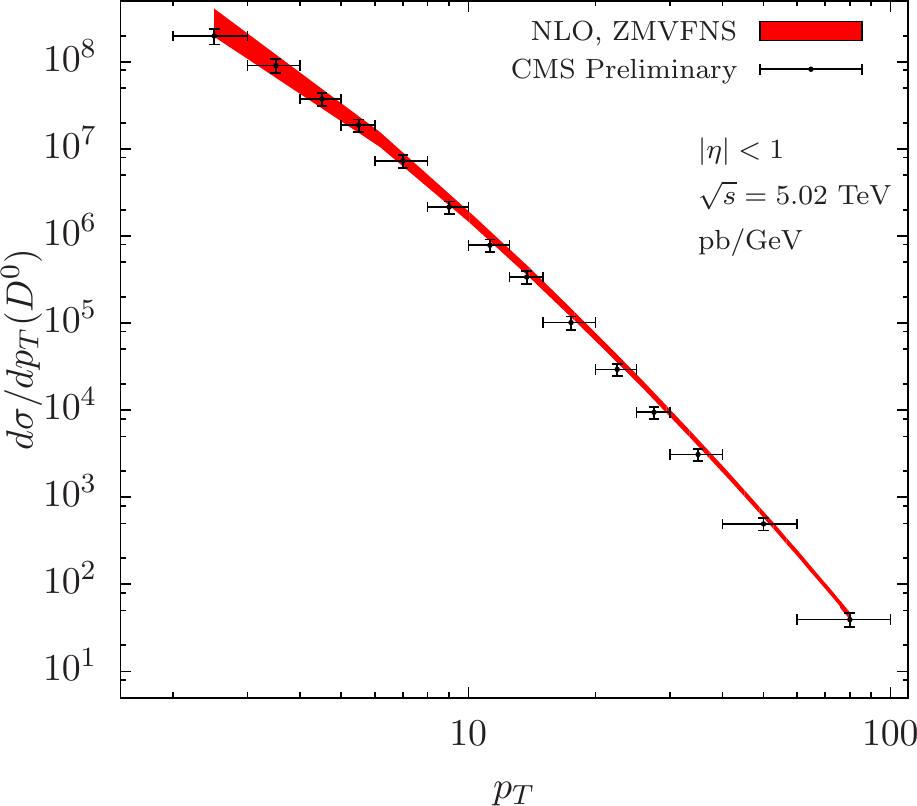} 
\end{center}
\vspace*{1.cm}
\caption{\label{fig:ppDX} The production cross sections for $pp\to D^{*\pm}X$ at $\sqrt{s}=7$~TeV (left) and for $pp\to D^0X$ at $\sqrt{s}=5.02$~TeV (right). The data was taken by the ATLAS collaboration~\cite{Aad:2015zix} for $D^{*\pm}$ and by the CMS collaboration~\cite{CMS:2016nrh} for $D^0$. Statistical and systematic errors are added in quadrature. The red curve is calculated within the ZM-VFNS scheme using the fragmentation functions of~\cite{Kneesch:2007ey,Kniehl:2009ar,Kniehl:2009mh,Kniehl:2012ti}. The band is obtained by varying $\mu_{R,F}$ by a factor of 2 around their central values of $\mu_{R,F}=m_T=\sqrt{p_T^2+m_c^2}$ and by taking the envelope.}
\end{figure*}

We use the $pp\to HX$ NLO framework developed in~\cite{Aversa:1988vb,Jager:2002xm} and typically applied for the production of light hadrons~\cite{Kretzer:2000yf,deFlorian:2007aj,deFlorian:2007ekg,Hirai:2007cx,Albino:2008fy,deFlorian:2014xna,Anderle:2015lqa,Epele:2016gup,Sato:2016wqj}. The double differential cross section can be written in the following way
\bea
\label{eq:sighadX}
\frac{d\sigma^{H}_{pp}}{dp_Td\eta}  = & \frac{2 p_T}{s}\sum_{a,b,c}\int_{x_a^{\rm min}}^1\f{dx_a}{x_a}f_a(x_a,\mu)\int_{x_b^{\rm min}}^1\f{dx_b}{x_b} f_b(x_b,\mu) 
\nnu
&\times \int^1_{z_c^{\rm min}} \frac{dz_c}{z_c^2}\frac{d\hat\sigma^c_{ab}(\hat s,\hat p_T,\hat \eta,\mu)}{dvdz}D_c^H(z_c,\mu),
\eea
where $\sum_{a,b,c}$ stands for a sum over all the parton flavors including light and heavy quarks and gluons, and $s$, $p_T$ and $\eta$ correspond to the center of mass energy, the hadron transverse momentum and hadron rapidity, respectively. Moreover, $f_{a,b}(x_{a,b},\mu)$ are the parton distribution functions for the two incoming protons. The hard functions $d\hat\sigma_{ab}^c(\hat s,\hat p_T,\hat\eta,\mu)$ are functions of the corresponding variables at the parton level: the partonic CM energy $\hat s=x_ax_bs$, the partonic transverse momentum $\hat p_T=p_T/z_c$ and the partonic rapidity $\hat\eta=\eta-\ln(x_a/x_b)/2$. The kinematical variables $v,z$ can be written in terms of these partonic variables
\bea
v=1-\f{2\hat p_T}{\sqrt{\hat s}}e^{-\hat\eta}, \qquad z=\f{2\hat p_T}{\sqrt{s}}\cosh\hat\eta \, .
\label{eq:vz}
\eea
Up to one loop order, the hard functions take the form
\bea
\frac{d\hat\sigma_{ab}^c}{dvdz} = \frac{d\hat\sigma_{ab}^{c,(0)}}{dv}\delta(1-z)+\frac{\alpha_s(\mu)}{2\pi} \frac{d\hat\sigma_{ab}^{c,(1)}}{dvdz}.
\eea
The integration limits in~(\ref{eq:sighadX}) are customarily written in terms of the hadronic variables $V,Z$,
\be
V=1-\f{2p_T}{\sqrt{s}}e^{-\eta}, \qquad Z=\f{2 p_T}{s}\cosh\eta \, ,
\ee
and are given by
\be
x_a^{\rm min}=1-\f{1-Z}{V},\quad x_b^{\rm min}=\f{1-V}{1+(1-V-Z)/x_a},\quad z_c^{\rm min}=\f{1-V}{x_b}-\f{1-V-Z}{x_a} \, .
\ee
On the other hand, $D_c^H(z_c,\mu)$ are the heavy meson fragmentation functions. For charmed mesons, we use the fragmentation functions of~\cite{Kneesch:2007ey,Kniehl:2009ar,Kniehl:2009mh,Kniehl:2012ti}, whereas for $B$-mesons we use the ones from~\cite{Kniehl:2008zza,Kniehl:2011bk,Kniehl:2015fla}. For $D$-mesons, the FFs are provided within the ZM-VFNS as well as in the General Mass Variable Flavor Number Scheme (GM-VFNS). For $B$-mesons, the FFs are only extracted using the GM-VFNS scheme. In the GM-VFNS, power corrections of the form $m_{c,b}^2/p_T^2$ are kept in the hard-scattering coefficients. However, the numerical relevance of these terms is small for sufficiently large $p_T$. See for example~\cite{Suzuki:1977km,Bowler:1981sb,Peterson:1982ak,Nason:1989zy,Nason:1993xx,Braaten:1994bz,Cacciari:1998it,Cacciari:2001td,Bauer:2013bza,Nejad:2013vsa,Fickinger:2016rfd} for other sets and possible approaches to heavy meson fragmentation functions. 

We start by presenting a comparison of NLO results in the ZM-VFNS with $D$-meson data taken by the ATLAS and CMS collaborations in Fig.~\ref{fig:ppDX}. In Fig.~\ref{fig:ppDX} (left), we show the ZM-VFNS results for $D^{*\pm}$ mesons at a CM energy of $\sqrt{s}=7$~TeV as a function of the transverse momentum. The rapidity is integrated over an interval of $|\eta|<2.1$. Note that here $D^{*\pm}$ does not correspond to the average but to the sum $D^{*\pm}=D^{*+}+D^{*-}$. Following~\cite{Kneesch:2007ey,Kniehl:2009ar,Kniehl:2009mh,Kniehl:2012ti}, we choose $\mu_{R,F}=m_T=\sqrt{p_T^2+m_c^2}$ as the central values for the renormalization and factorization scales. The band is obtained by varying $\mu_{R,F}$ independently around their central values by a factor of 2 and by taking the envelope. Along with the theoretical calculation, we show the ATLAS data of~\cite{Aad:2015zix}. Throughout this section, we always show the combined statistical and systematic errors added in quadrature. Analogously, in Fig.~\ref{fig:ppDX} (right), we show the results for $D^0$ production at $\sqrt{s}=5.02$~TeV with $|\eta|<1$ comparing to preliminary data from CMS~\cite{CMS:2016nrh}. Keeping in mind that the $D$-meson fragmentation functions of~\cite{Kneesch:2007ey,Kniehl:2009ar,Kniehl:2009mh,Kniehl:2012ti} are fitted to $e^+e^-$ data only, we find that the agreement between theory and data is indeed remarkably good. In addition, we would like to emphasize that the agreement between the NLO calculation within the ZM-VFNS and the data is still good even at relatively low values of $p_T$ of a few GeV.

\begin {figure*}[t]
\begin{center}
\vspace*{10mm}
\includegraphics[width=0.4\textwidth,trim=1cm 2cm 1cm 1cm ]{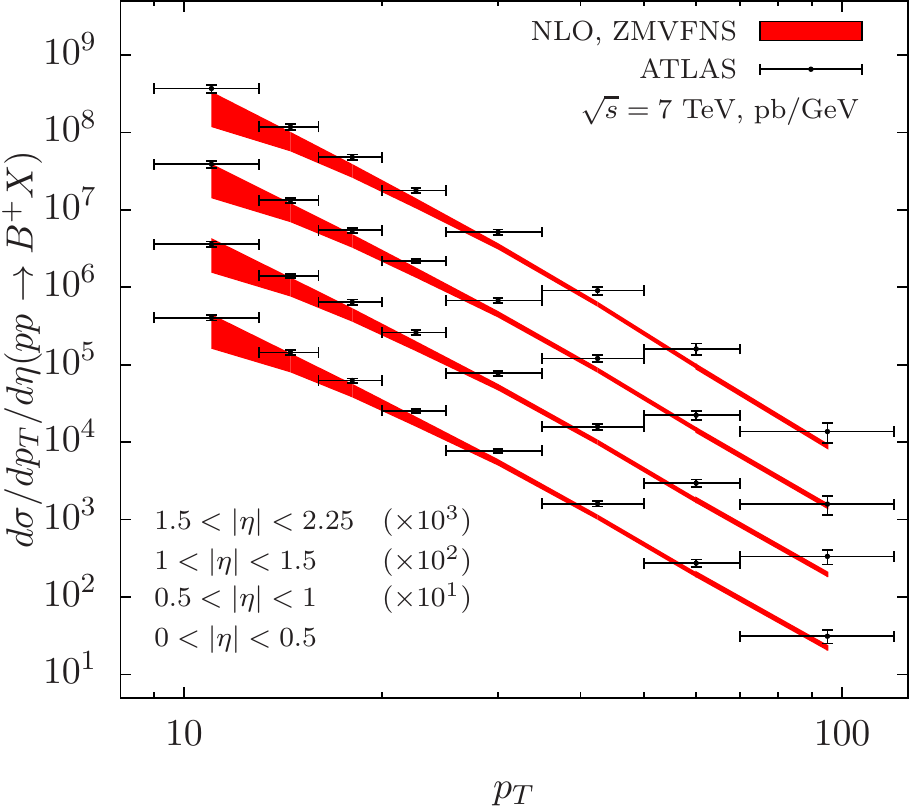} 
\hspace*{2cm}
\includegraphics[width=0.4\textwidth,trim=1cm 2cm 1cm 1cm ]{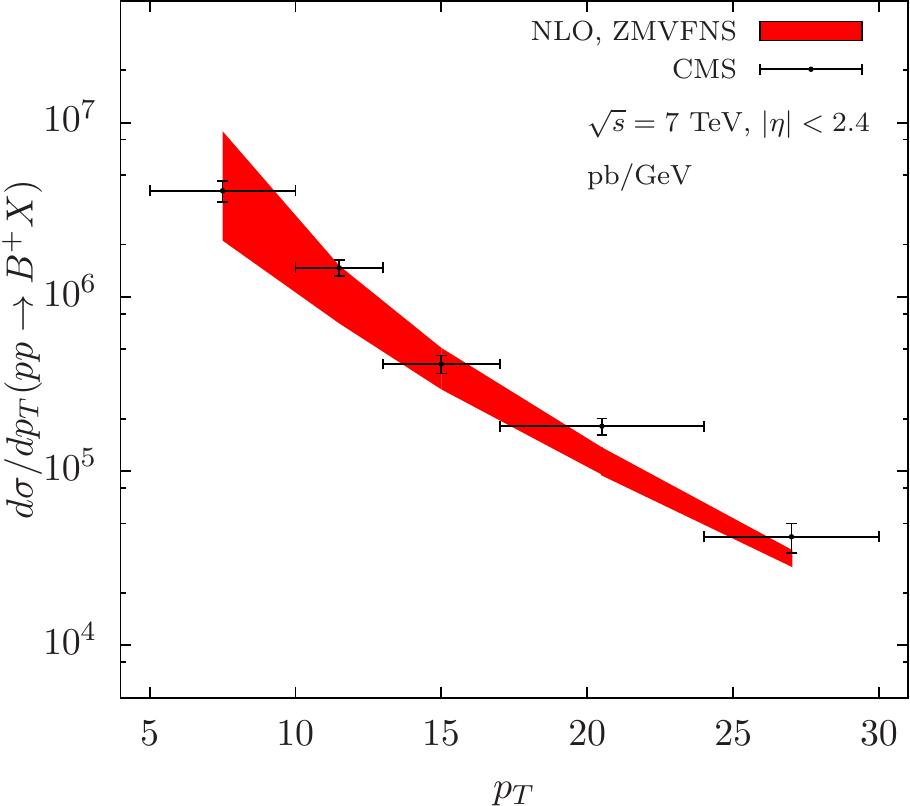} 
\end{center}
\vspace*{1.cm}
\caption{\label{fig:ppBX} The $pp\to B^+X$ production cross section for $\sqrt{s}=7$~TeV within the ZM-VFNS~\cite{Kniehl:2008zza,Kniehl:2011bk,Kniehl:2015fla} in comparison to data from ATLAS~\cite{ATLAS:2013cia} (left) and CMS~\cite{Khachatryan:2011mk} (right). The ATLAS data is presented for four different rapidity intervals in the range of $\eta=0-2.25$ whereas the CMS results is for $|\eta|<2.4$. Statistical and systematic errors are added in quadrature. The NLO calculation is shown in red where the factorization and renormalization scales are chosen as $\mu_{R,F}=m_T=\sqrt{p_T^2+m_b^2}$. The band is obtained by varying $\mu_{R,F}$ independently by a factor of 2 around their central values and by taking the envelope.}
\end{figure*}

Similarly, in Fig.~\ref{fig:ppBX}, we show analogous comparisons for $B$-mesons. We choose to only show two exemplary comparisons of the NLO calculation in the ZM-VFNS and inclusive $pp\to B^+X$ data from ATLAS~\cite{ATLAS:2013cia} (left) and CMS~\cite{Khachatryan:2011mk} (right). For both data sets, $B^+$ mesons are identified via the exclusive decay channel $B^+\to J/\psi K^+\to \mu^+\mu^-K^+$. The corresponding multiplicative branching fractions are taken into account in our calculations. The CMS data is integrated over $|\eta|<2.4$, whereas the ATLAS data is presented for four different rapidity intervals in the range of $\eta=0-2.25$. Both data sets were taken at a CM energy of $\sqrt{s}=7$~TeV. The default scale for the NLO calculation is now $\mu_{R,F}=m_T=\sqrt{p_T^2+m_b^2}$ following~\cite{Kniehl:2008zza,Kniehl:2011bk,Kniehl:2015fla}. Again, the band is obtained by varying $\mu_{R,F}$ independently around their default choice by a factor of 2 and by taking the envelope. Similar to the inclusive $D$-meson production, the agreement between theory and data is remarkably good even down to relatively low $p_T$.

\begin {figure*}[t]
\begin{center}
\vspace*{10mm}
\includegraphics[width=0.4\textwidth,trim=1cm 2cm 1cm 1cm ]{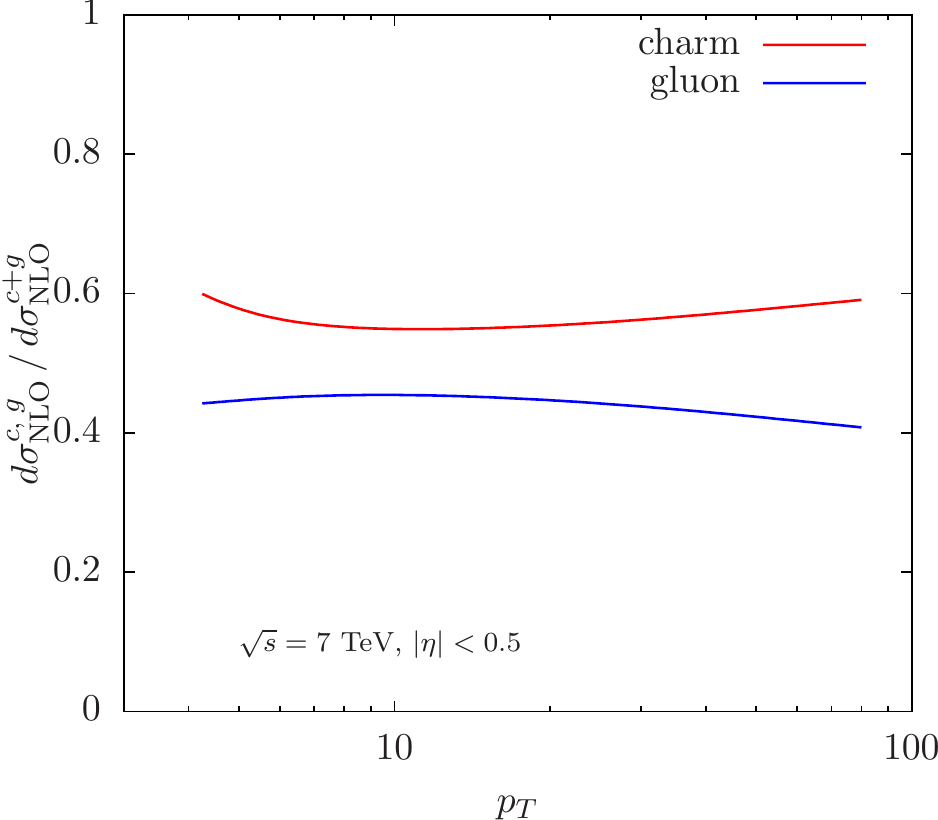} 
\hspace*{2cm}
\includegraphics[width=0.4\textwidth,trim=1cm 2cm 1cm 1cm ]{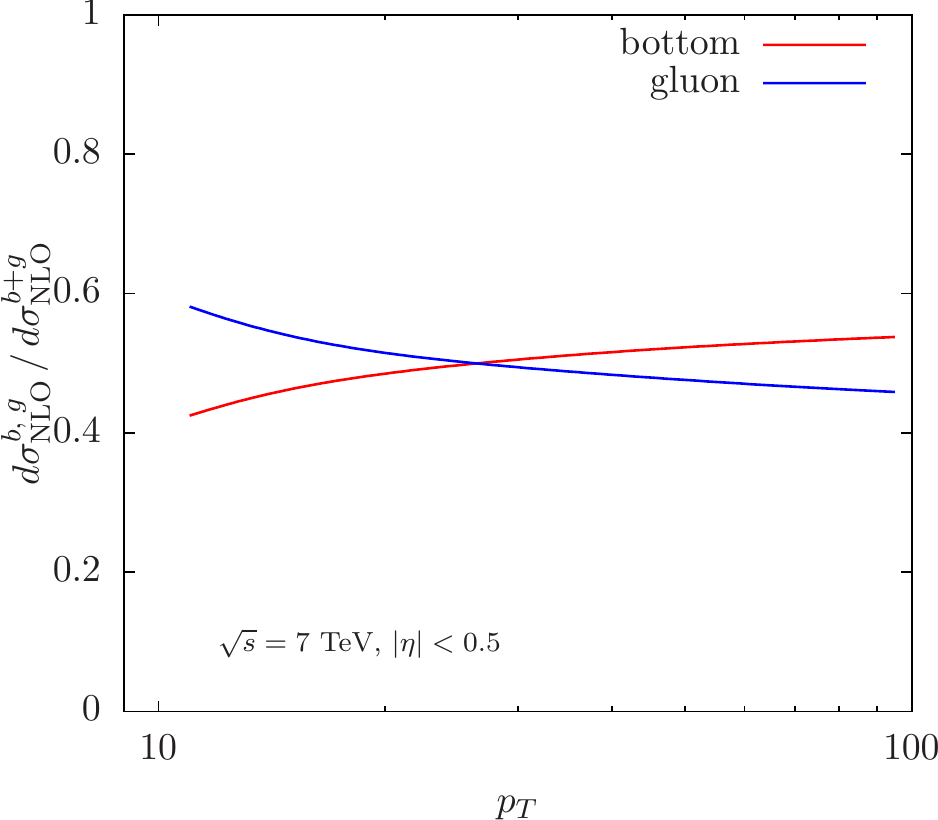} 
\end{center}
\vspace*{1.cm}
\caption{\label{fig:cbg} The percentage contribution of the heavy quark and gluon fragmentation processes to inclusive $D$-meson (left) and $B$-meson (right) production at NLO for $\sqrt{s}=7$~TeV. The heavy quark contribution (charm, bottom) is shown in red and the gluon is shown in blue. We use the fragmentation functions of~\cite{Kneesch:2007ey,Kniehl:2009ar,Kniehl:2009mh,Kniehl:2012ti} (left) and~\cite{Kniehl:2008zza,Kniehl:2011bk,Kniehl:2015fla} (right) within the ZM-VFNS and we have as usual $\mu_{R,F}=m_T=\sqrt{p_T^2+m_{c,b}^2}$.}
\end{figure*}

We would like to point out an important difference to several earlier calculations in the literature. Often the heavy meson production is calculated differently and only the modification of the heavy-quark-to-heavy-meson fragmentation process is taken into account in Pb+Pb collisions. However, in the ZM-VFNS, there is a large gluon-to-heavy-meson contribution, even though the gluon-to-heavy-meson fragmentation function itself is much smaller than the corresponding heavy-quark-to-heavy-meson FF. The smallness of the gluon FF itself is compensated by the large gluon production cross section in proton-proton collisions at high CM energies. In fact, as illustrated in Fig.~\ref{fig:cbg}, the gluon-to-heavy-meson contribution (shown in blue) is of the order of 50$\%$ for both $D$- (left) and $B$-meson (right) production in $pp$ collisions at $\sqrt{s} = 7$ TeV. The percentage contribution of the heavy-quark-to-heavy-meson fragmentation process is shown in red for both $D$- and $B$-meson production. We note that the light-quark-to-heavy-meson fragmentation contribution turns out to have a marginal effect only. 
We would like to stress again that heavy meson FFs are generally extracted from $e^+e^-$ data only. In this case, the gluon-to-heavy-meson FF only enters at the one-loop level and through evolution effects. This leads to the fact that the gluon fragmentation function is relatively poorly constrained from $e^+e^-$ alone. In the future, it will be very helpful to obtain heavy meson fragmentation functions within a global analysis including in particular $pp\to HX$ data as it is customarily done for light hadrons~\cite{deFlorian:2007aj,deFlorian:2007ekg,deFlorian:2014xna}. In addition, including in-jet fragmentation data $pp\to(\mathrm{jet}H)X$~\cite{Aad:2011td}, an observable for which a new theory framework was recently developed~\cite{Procura:2009vm,Jain:2011xz,Arleo:2013tya,Kaufmann:2015hma,Chien:2015ctp,Kaufmann:2016nux,Kang:2016mcy,Kang:2016ehg,Bain:2016clc}, is expected to lead to great improvements of the corresponding global fits. 

Whether a gluon-to-heavy-meson fragmentation function is included in the calculation or not is especially relevant for the in-medium calculation. The energy loss of heavy quarks and gluons in the medium is very different. Therefore, it is absolutely crucial to understand how heavy mesons are formed in order to obtain a reliable quantitative understanding of their suppression in heavy-ion collisions. In fact, not only the relative percentage of gluon and heavy quark fragmentation as shown in Fig.~\ref{fig:cbg} is important but also the exact shape of the fragmentation functions is relevant. The inclusive hadron spectra $pp\to HX$ are relatively well described by currently available sets of fragmentation functions as shown above. However, for example, the disagreement between theory and data for heavy mesons measured inside jets clearly shows that the currently available fragmentation functions are still not well enough understood so far, see~\cite{Chien:2015ctp}. Performing new global fits is beyond the scope of this work but we are planning to addressed this issue in future publications.

\subsection{Suppression of $D$- and $B$-mesons in Pb+Pb collisions at the LHC}

\begin {figure*}[t]
\begin{center}
\vspace*{10mm}
\includegraphics[width=0.4\textwidth,trim=1cm 2cm 1cm 1cm ]{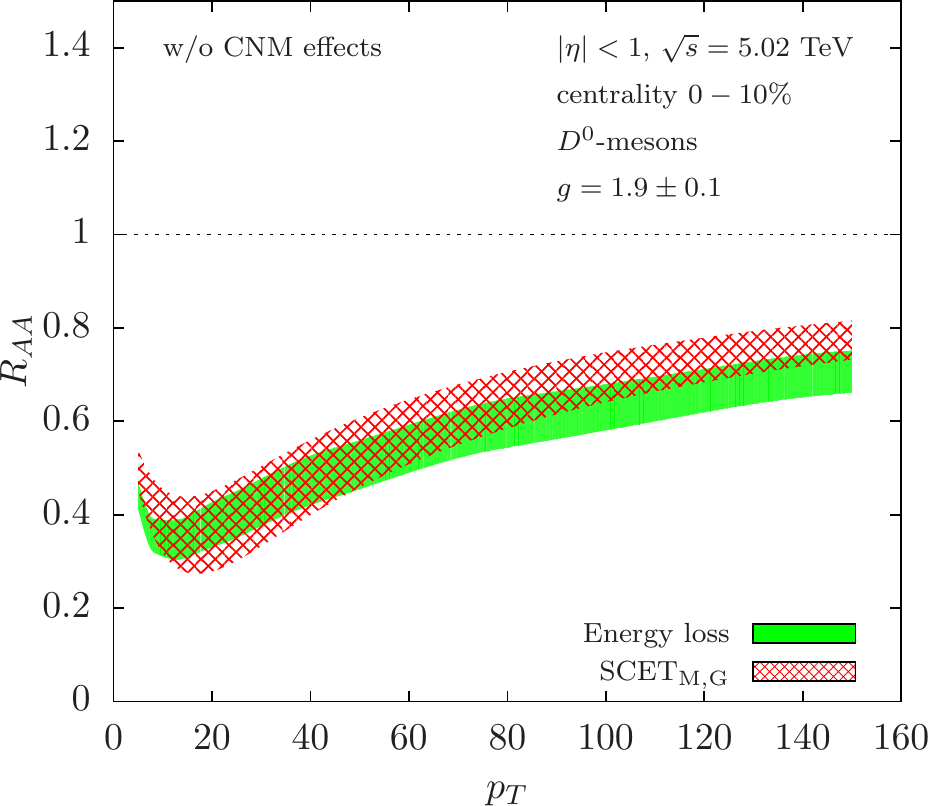} 
\hspace*{2cm}
\includegraphics[width=0.4\textwidth,trim=1cm 2cm 1cm 1cm ]{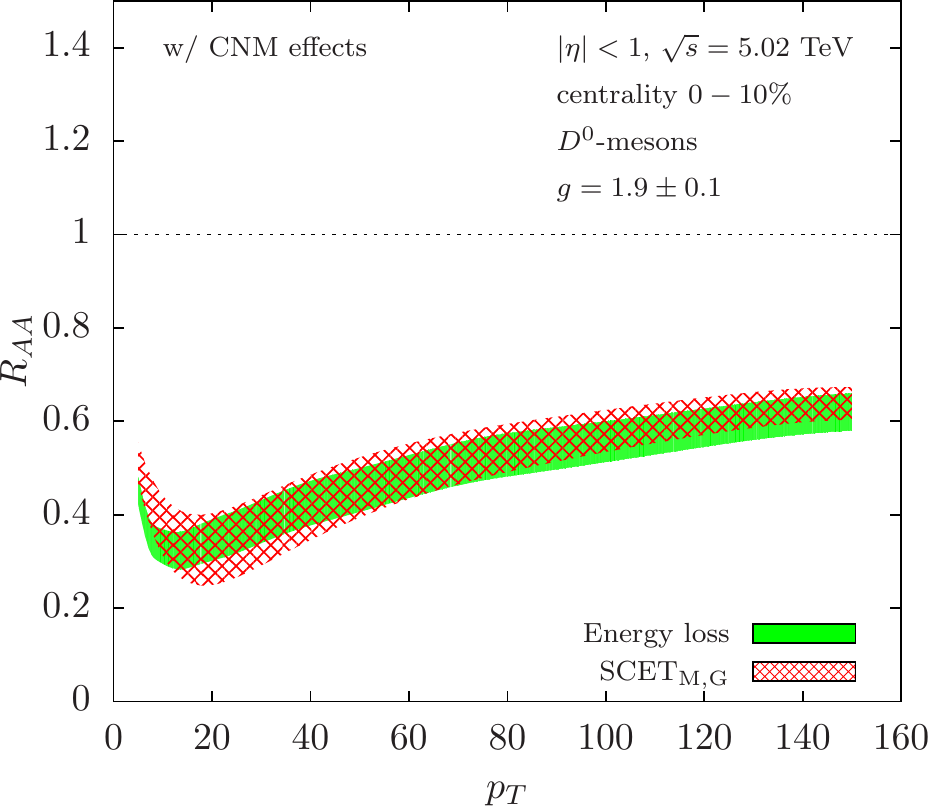} 
\end{center}
\vspace*{1.cm}
\caption{\label{fig:RAA-D} The nuclear modification factor $R_{AA}$ for $D^0$ meson production as a function of the transverse momentum $p_T$. We show the result obtained within the traditional approach to energy loss (green band) as well as the new results based on SCET$_{\mathrm{M,G}}$ (hatched red band). We choose a CM energy of $\sqrt{s_\mathrm{NN}}=5.02$~TeV, a rapidity interval of $0<|\eta|<1$ and the results are for central collisions with centrality $0-10\%$. In addition, we choose the coupling strength as $g=1.9\pm 0.1$. The results for $R_{AA}$ are presented without CNM effects (left) and with CNM effects (right). See text for more details.}
\end{figure*}

In this section, we present the results for the nuclear modification factor $R_{AA}$ defined as
\be
R_{AA}=\f{d\sigma^H_{\mathrm{PbPb}}/d\eta dp_T}{\braket{N_{\mathrm{bin}}} d\sigma^H_{pp}/d\eta dp_T} \, .
\ee
Here, $\braket{N_{\mathrm{bin}}}$ is the average number of binary nucleon-nucleon collisions for a given centrality and $d\sigma^H_{pp}/d\eta dp_T$, $d\sigma^H_{\mathrm{PbPb}}/d\eta dp_T$ are the double differential cross sections for inclusive heavy meson $H$ production in proton-proton and Pb+Pb collisions respectively. The cross section for proton-proton collision was given in Eq.~(\ref{eq:sighadX}) and its modification for Pb+Pb collisions was discussed in the previous section, cf.~Eq.~\eqref{eq:AA}. Our calculations depend on one parameter and the result of initial-state effects. Firstly, there is the coupling constant $g$ that describes how strongly the hard partons couple to the QCD medium. As in several earlier publications~\cite{Kang:2014xsa,Chien:2015vja}, we choose this parameter around $g\approx 2$. When presenting numerical results for the nuclear modification factor $R_{AA}$, we typically vary this parameter around its central value by $\pm 0.1$ and plot the obtained band. Eventually, the coupling strength $g$ will have to be constrained by comparing to data. The second set of effects are Cold Nuclear Matter (CNM) effects, which happen before the formation of the QGP. These include isospin effects, coherent power corrections for heavy quarks~\cite{Qiu:2004qk,Vitev:2006bi}, the Cronin effect~\cite{Accardi:2002ik} and cold nuclear matter energy loss~\cite{Vitev:2007ve,Kang:2015mta,Ovanesyan:2015dop}. By implementing isospin effects we take into account that the Pb nucleus is made up of both protons and neutrons. As discussed above and illustrated in Fig.~\ref{fig:cbg}, we find that for heavy meson production only heavy quark and gluon fragmentation functions turn out to be relevant. These processes are isospin symmetric. Therefore, for all our results presented here, the isospin effects are very small. The effect of power corrections is limited to small $ p_T$ and does not affect the ZM-VFNS region of applicability.  The Cronin effect and CNM energy loss effects can partly be constrained by p+Pb collisions for example. It is clear that there is always a non-trivial interplay between the value of the coupling strength $g$ and possible CNM effects. To some extent a variation of, say, $g$ can be absorbed by changing the strength of CNM effects. That being said, we would like to point out that the two effects are not completely interchangeable and it is indeed possible to constrain both effects from precision data in Pb+Pb collisions when observables are carefully selected, see also~\cite{Chien:2015vja}.

\begin {figure*}[t]
\begin{center}
\vspace*{10mm}
\includegraphics[width=0.4\textwidth,trim=1cm 2cm 1cm 1cm ]{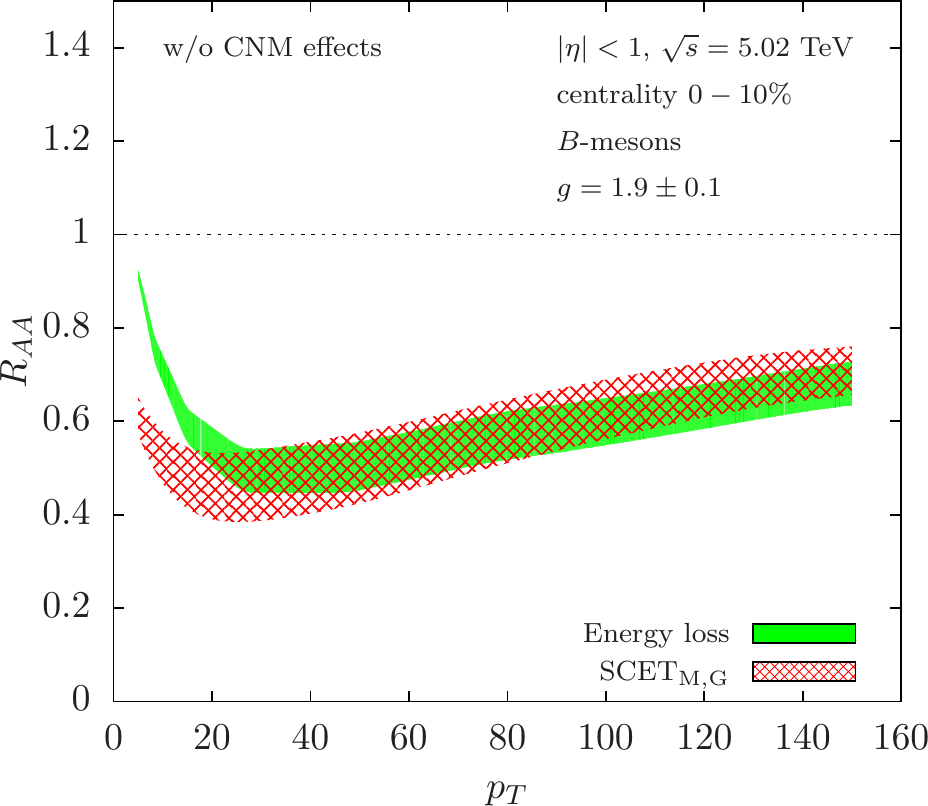} 
\hspace*{2cm}
\includegraphics[width=0.4\textwidth,trim=1cm 2cm 1cm 1cm ]{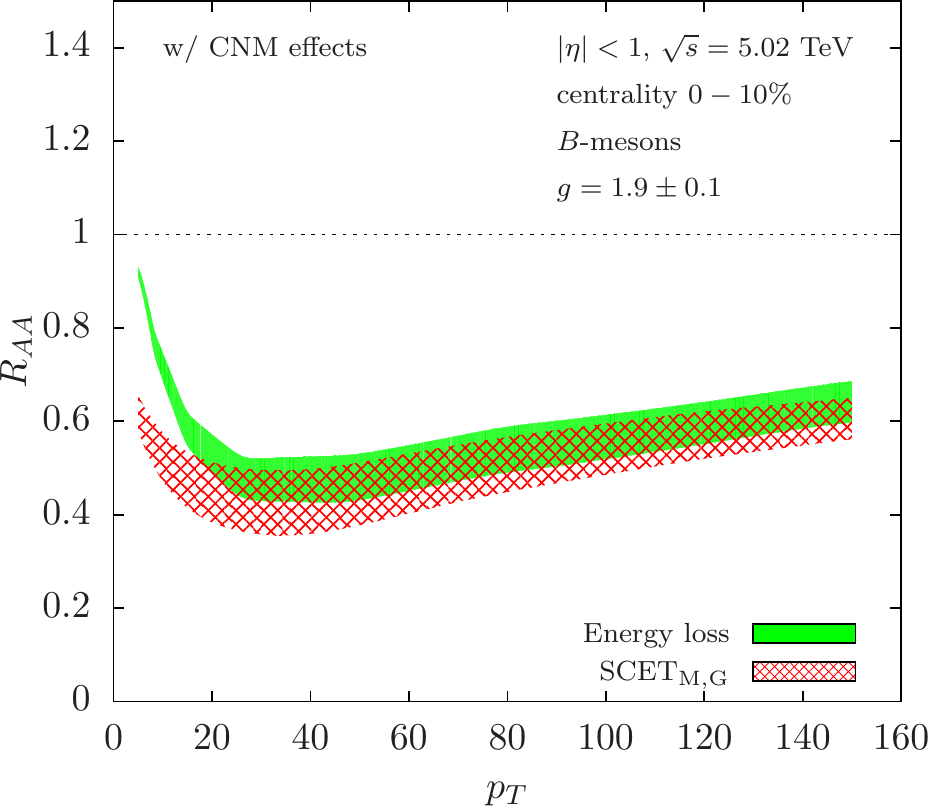} 
\end{center}
\vspace*{1.cm}
\caption{\label{fig:RAA-B} Same as Fig.~\ref{fig:RAA-D} but for $B^+$ mesons.}
\end{figure*}

We start by presenting results for the $R_{AA}$ of $D^0$-mesons using the new framework of SCET$_{\mathrm{M,G}}$ (hatched red band) in Fig.~\ref{fig:RAA-D}. In addition, we present results obtained within the traditional approach to parton energy loss (green band). We choose a CM energy of $\sqrt{s_\mathrm{NN}}=5.02$~TeV and integrate over the rapidity interval $|\eta|<1$. The charm mass is chosen as $m_c=1.3$~GeV. The results are presented for central collisions with centrality $0-10\%$. The bands are obtained by varying the coupling strength around its central value $g=1.9\pm 0.1$. On the left (right) hand side of Fig.~\ref{fig:RAA-D}, we show the results without (with) CNM effects. It can be seen that the CNM effects can affect the $R_{AA}$ both at low and high-$p_T$. As it turns out, they lead to a rise at relatively low-$p_T$, whereas a suppression in the high-$p_T$ region is observed. We find that both the SCET$_{\mathrm{M,G}}$ results and the results based on traditional parton energy loss are quite similar in the large-$p_T$ region. However, at low-$p_T$, the two results differ significantly. We would like to point out that the difference between the two results is not entirely due to the different theoretical approaches to the in-medium interaction SCET$_{\mathrm{M,G}}$ vs. traditional parton energy loss. For both $D$- and $B$-meson (see Fig.~\ref{fig:RAA-B}) production within SCET$_{\mathrm{M,G}}$, we use modern fits of fragmentation functions that include both heavy quark and gluon fragmentation. Instead, the results presented here using the traditional picture of parton energy loss are calculated using only heavy quark fragmentation functions based on a model calculation~\cite{Braaten:1994bz}, as it is conventionally done in the literature, see e.g. \cite{Sharma:2009hn}. The different choice of fragmentation functions can lead to a very different result for the $R_{AA}$ in particular at low $p_T$. We discuss this point in more detail below.

In Fig.~\ref{fig:RAA-B}, we present analogous results for the nuclear modification factor for $B^+$ meson production. The bottom mass is chosen as $m_b=4.5$~GeV. By comparing Figs.~\ref{fig:RAA-D} and~\ref{fig:RAA-B} one notices that there is indeed a difference of the $R_{AA}$ between $D^0$- and $B^+$-meson suppression independent of the approach (medium-induced splitting only). Firstly, this is in part due to the different fragmentation functions. Secondly, the different masses for charm and bottom quarks can affect the $R_{AA}$ even at relatively large $p_T$. As can be seen from Fig.~\ref{fig:RAA-B}, the difference between SCET$_{\mathrm{M,G}}$ based results and the results from traditional parton energy loss differ more significantly at low-$p_T$ than it is the case for $D^0$-mesons. The large difference between the two approaches at low $p_T$ is mainly due to the fact that for the traditional parton energy loss calculation we only take into account heavy quark fragmentation functions.

\begin {figure*}[t]
\begin{center}
\vspace*{10mm}
\includegraphics[width=0.4\textwidth,trim=1cm 2cm 1cm 0.5cm ]{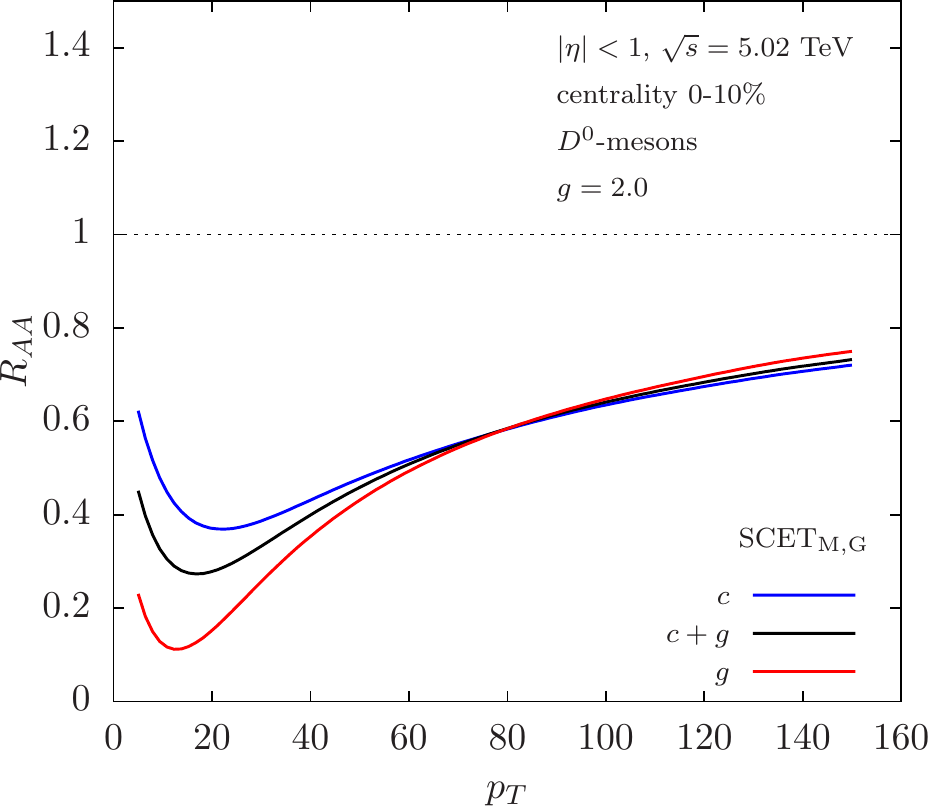} 
\hspace*{2cm}
\includegraphics[width=0.4\textwidth,trim=1cm 2cm 1cm 0.5cm ]{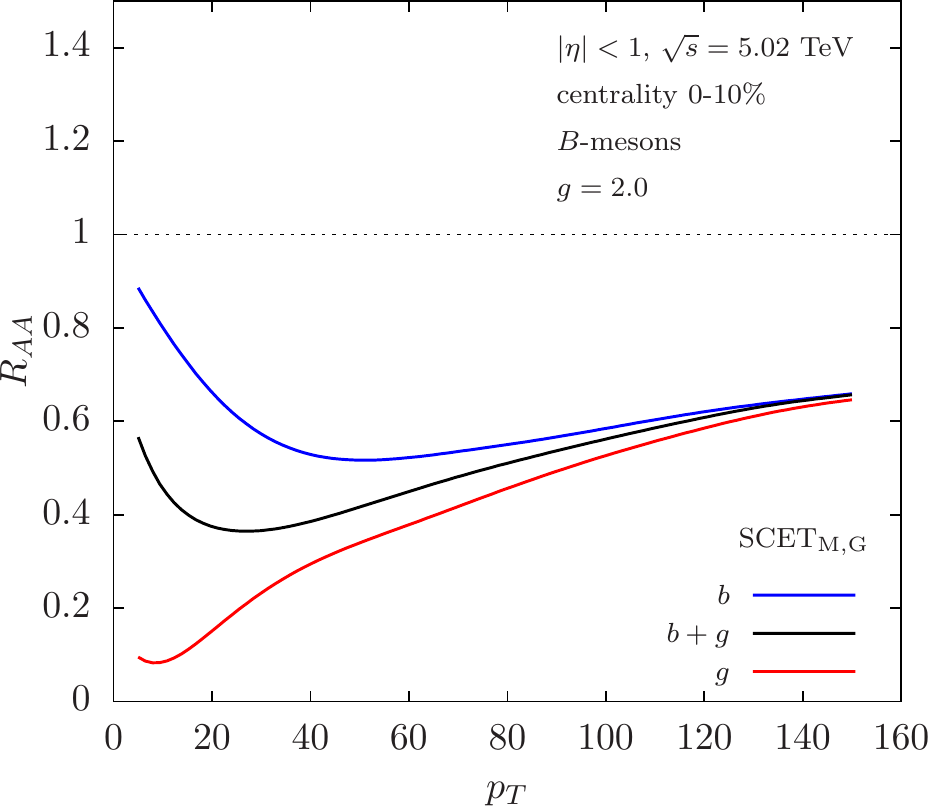} 
\end{center}
\vspace*{1.cm}
\caption{\label{fig:cbgRAA} Nuclear modification factor $R_{AA}$ for $D^0$ mesons (left) and $B$-mesons (right). As an example, we choose $\sqrt{s_\mathrm{NN}}=5.02$~TeV, $|\eta|<1$ and $0-10\%$ centrality. In black the standard $R_{AA}$ is shown as in Figs.~\ref{fig:RAA-D} and~\ref{fig:RAA-B}. In blue, we show the suppression for the ``heavy quark only'' case. This result is obtained by calculating both the proton-proton baseline as well as the medium effects with only charm (left) and bottom quark fragmentation functions (right). In red, the analogous result is shown for the ``gluon only'' case.}
\end{figure*}

As already pointed out in the previous section, it is of great relevance to understand the relative contributions of heavy quark and gluon fragmentation to the heavy meson production cross sections~\cite{Huang:2013vaa,Huang:2015mva}. Fig.~\ref{fig:cbgRAA} illustrates the implications for the obtained nuclear modification factor $R_{AA}$. In black, we show the combined calculation for the suppression of $D^0$ (left) and $B^+$ mesons (right) in heavy-ion collisions as before in Figs.~\ref{fig:RAA-D} and~\ref{fig:RAA-B}. In blue, the suppression is shown for the ``heavy quark only'' case. These results are obtained by calculating both the proton-proton baseline as well as the in-medium effects with only charm quark (left) and bottom quark (right) fragmentation functions. In red, we show the analogous result for the ``gluon only'' case. While the suppression for heavy quarks and gluons is similar in the high $p_T$ region, it turns out that their suppression is very different in the low $p_T$ region. The difference between fragmenting gluons and heavy quarks is more pronounced for the heavier $B^+$-mesons. The very different suppression rates for heavy quarks and gluons can lead to a significantly different picture of how the  QCD medium affects open heavy flavor. Besides these important differences, there are two main sources of uncertainties at low $p_T$. Firstly, the gluon-to-heavy-meson fragmentation function is still relatively poorly constrained. This concerns both the exact functional form as well as the total contribution to the cross section of gluons as discussed above and illustrated in Fig.~\ref{fig:cbg}. Therefore, we would like to stress that a more reliable picture of the in-medium interactions requires further improvements already at the level of the proton-proton baseline calculation. Secondly, in the low $p_T$ region higher order terms in the opacity series expansion are expected to play a more important role. In the future we plan to address these issues in order to systematically improve the current framework allowing an extension to lower values of $p_T$. In addition, other effects like collisional energy loss and dissociation are expected to play a role at low $p_T$ as well~\cite{Wicks:2005gt,Adil:2006ra}. However, given the currently remaining uncertainties at low $p_T$, it is clear that further improvements are needed before making any definitive statements about where exactly other effects are relevant or even dominate. We note that our conclusions here are different than in~\cite{Cao:2015kvb}, where the authors concluded that the energy loss for gluons that fragment into heavy mesons is small because gluons quickly split into heavy quark anti-quark pairs which then fragment into the observed heavy mesons. Instead, here we are motivated by QCD factorization saying that the actual hadronization is a long-distance effect which happens at much later time scales than the hard-scattering event. At least within the ZM-VFNS, gluon fragmentation has been put on an equal footing as the heavy-quark fragmentation function within the QCD factorization formalism, as can be seen clearly in Eqs.~\eqref{eq:AA} and \eqref{eq:sighadX}. In this sense, our approach is in direct analogy to light charged hadron production.  

We would like to add that a unique opportunity to test and improve the current theoretical framework would be to measure and calculate the in-jet fragmentation of heavy mesons both in proton-proton and heavy-ion collisions~\cite{Aad:2011td,Chien:2015ctp}. Firstly, the poorly known gluon-to-heavy-meson fragmentation functions can be studied at a more differential level. Secondly, the in-jet fragmentation will allow to disentangle better the modification of the two main fragmentation contributions to heavy meson production.

\begin {figure*}[t]
\begin{center}
\vspace*{10mm}
\includegraphics[width=0.4\textwidth,trim=1cm 2cm 1cm 0.5cm ]{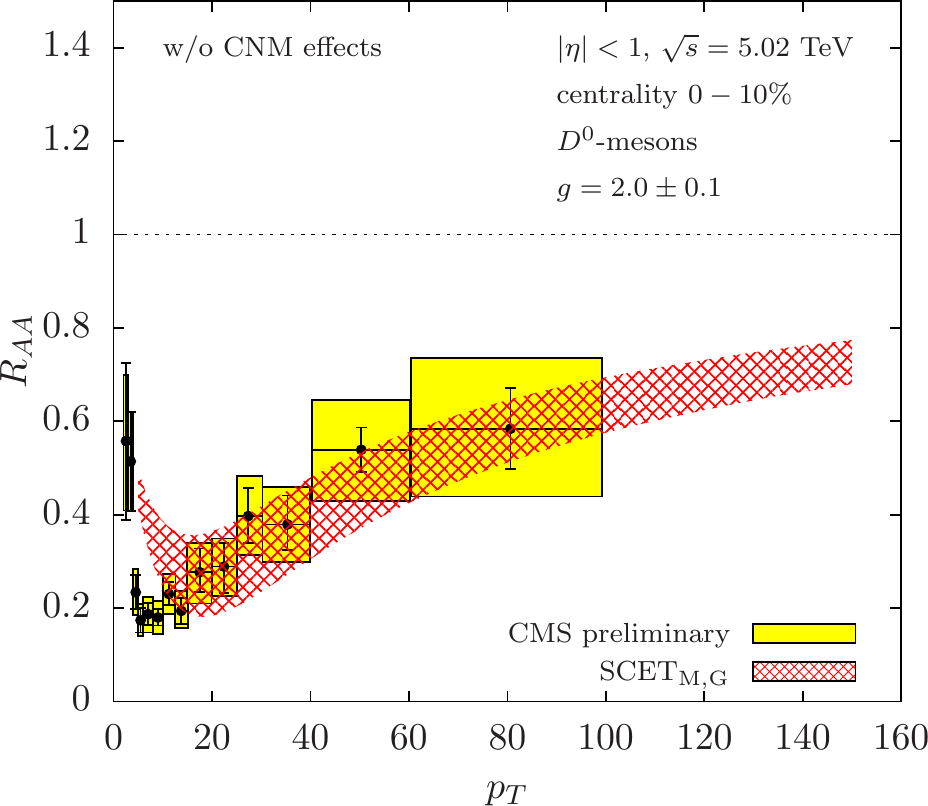} 
\hspace*{2cm}
\includegraphics[width=0.4\textwidth,trim=1cm 2cm 1cm 0.5cm ]{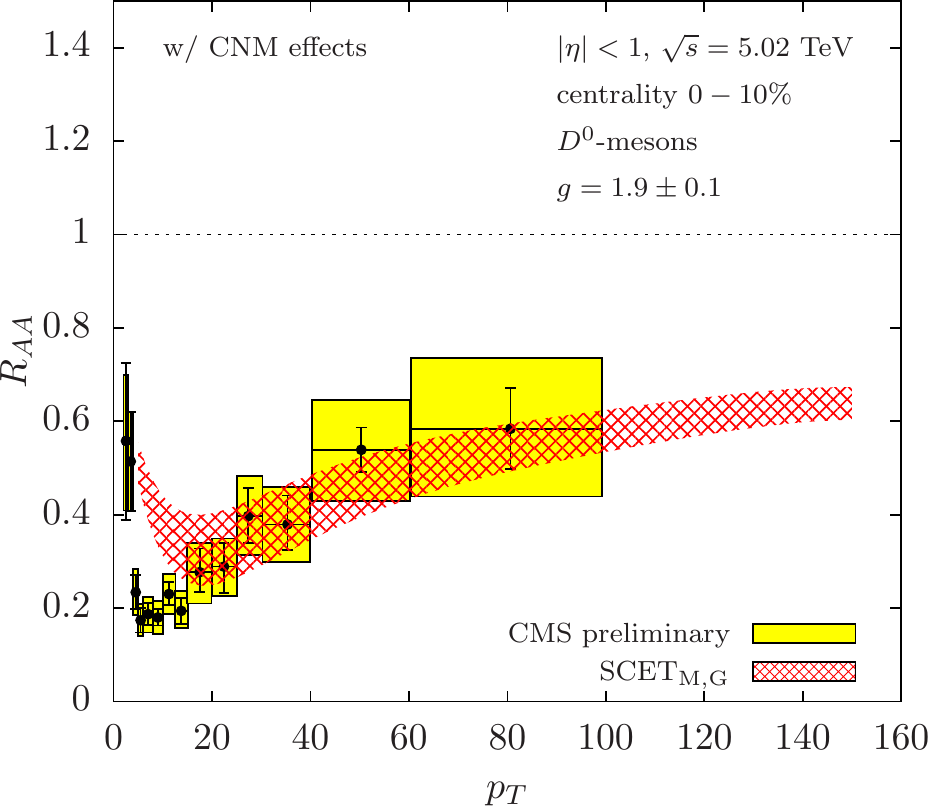} 
\end{center}
\vspace*{1.cm}
\caption{\label{fig:RAA-D-exp-CMS} Nuclear modification factor $R_{AA}$ for $D^0$ mesons without (left) and with (right) CNM effects in comparison to preliminary the CMS data of~\cite{CMS:2016nrh}. We have $\sqrt{s_\mathrm{NN}}=5.02$~TeV, $|\eta|<1$ and $0-10\%$ centrality. When CNM effects are (not) included, we choose the coupling strength as $g=1.9\pm 0.1$ ($g=2.0\pm 0.1$).}
\end{figure*}

Despite the remaining uncertainties at low $p_T$, we expect to have a reliable description of the in-medium effects as long as the transverse momentum of the observed heavy meson is sufficiently large $p_T\gtrsim 10$~GeV. We proceed by comparing our new SCET$_{\mathrm{M,G}}$ based calculations with currently available experimental data from CMS and ALICE as an example. In Fig.~\ref{fig:RAA-D-exp-CMS}, we present a comparison to the preliminary CMS data of~\cite{CMS:2016nrh} for the nuclear modification factor $R_{AA}$ for $D^0$-mesons. The data was taken for $\sqrt{s_\mathrm{NN}}=5.02$~TeV, $|\eta|<1$ and $0-10\%$ centrality. Note that for all the data presented in this section, we show the statistical errors as standard error bars and the systematic ones as yellow boxes. On the left hand side, we compare the data to our calculation without CNM effects and we choose the coupling strength as $g=2.0\pm 0.1$. Instead, on the right hand side, we include CNM effects as discussed above. As illustrated in Fig.~\ref{fig:RAA-D-exp-CMS}, it turns out that CNM effects lead to a larger suppression at high-$p_T$. Therefore, we choose a lower coupling strength $g=1.9\pm 0.1$ for the comparison to data. Again, we would like to point out that one effect can not be compensate entirely by the other one. Both calculations based on the newly derived SCET$_{\mathrm{M,G}}$ agree very well with the data in the expected $p_T$ region. While the result without CNM effects seems to agree slightly better with the data, one can not make any definitive statement given the experimental uncertainties.

\begin {figure*}[t]
\begin{center}
\vspace*{10mm}
\includegraphics[width=0.4\textwidth,trim=1cm 2cm 1cm 0.5cm ]{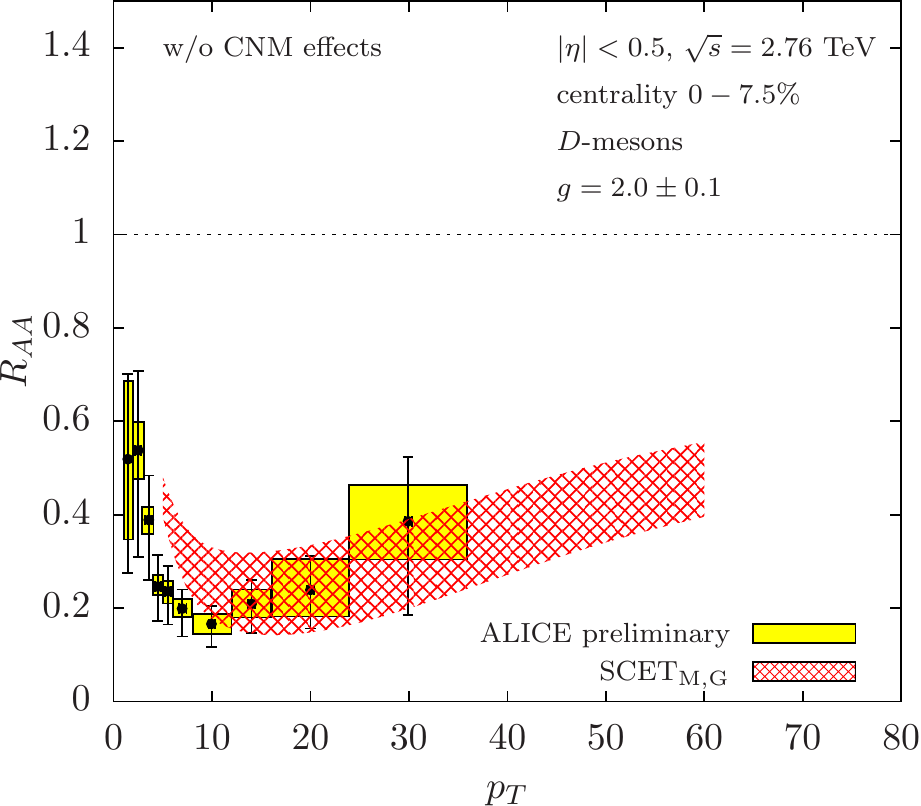} 
\hspace*{2cm}
\includegraphics[width=0.4\textwidth,trim=1cm 2cm 1cm 0.5cm ]{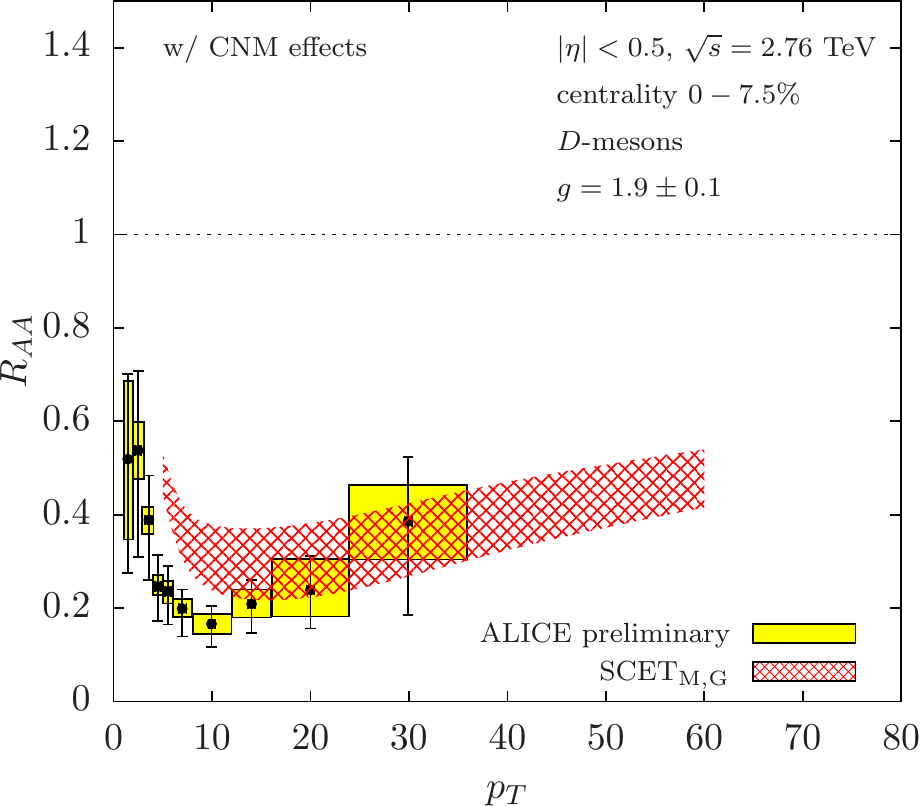} 
\end{center}
\vspace*{1.cm}
\caption{\label{fig:RAA-D-exp} Nuclear modification factor $R_{AA}$ for $D$-mesons ($D^0$, $D^+$ and $D^{*+}$ average) without (left) and with (right) CNM effects in comparison to preliminary ALICE data of~\cite{Grelli:2012yv}. We have $\sqrt{s_\mathrm{NN}}=2.76$~TeV, $|\eta|<0.5$ and $0-7.5\%$ centrality. When CNM effects are (not) included, we choose the coupling strength as $g=1.9\pm 0.1$ ($g=2.0\pm 0.1$).}
\end{figure*}

Next, we compare to ALICE data~\cite{Grelli:2012yv} at $\sqrt{s_\mathrm{NN}}=2.76$~TeV in Fig.~\ref{fig:RAA-D-exp}. The nuclear modification factor $R_{AA}$ is shown for $D$-mesons ($D^0$, $D^+$ and $D^{*+}$ average) with $|\eta|<0.5$ and $0-7.5\%$ centrality. Again we present our numerical results including CNM effects for $g=2.0\pm 0.1$ (left) and the results without CNM effects for $g=1.9\pm 0.1$ (right). We find that the data is well described by our calculations for $p_T\gtrsim 10$~GeV. Our result without CNM effects seems to agree slightly better with the data.

\begin {figure*}[t]
\begin{center}
\vspace*{10mm}
\includegraphics[width=0.4\textwidth,trim=1cm 2cm 1cm 0.5cm ]{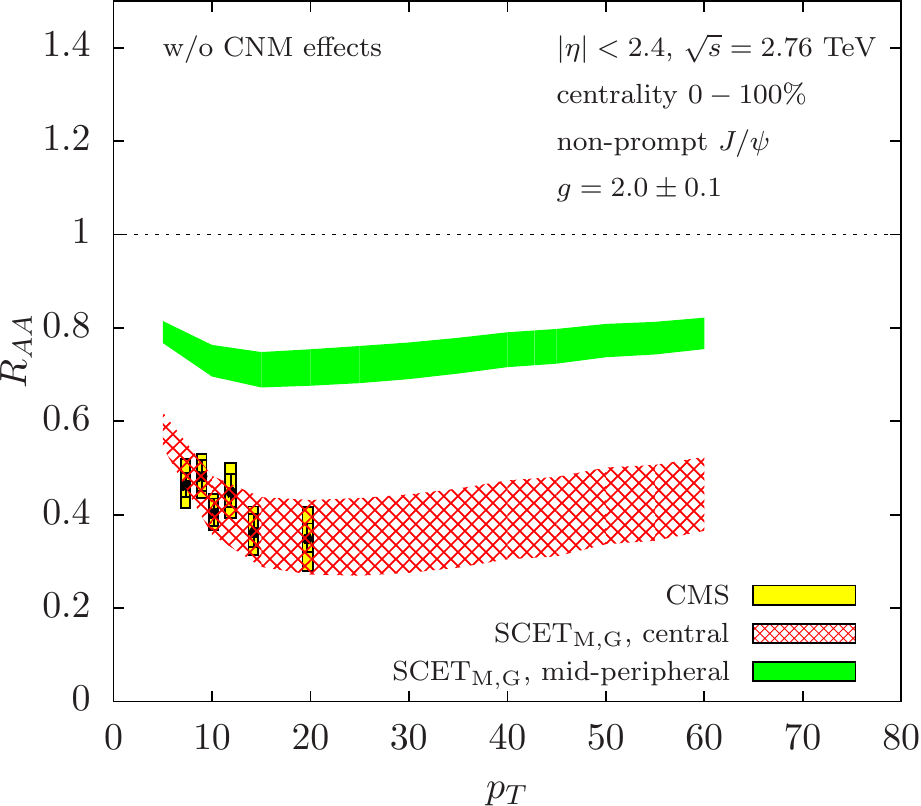} 
\hspace*{2cm}
\includegraphics[width=0.4\textwidth,trim=1cm 2cm 1cm 0.5cm ]{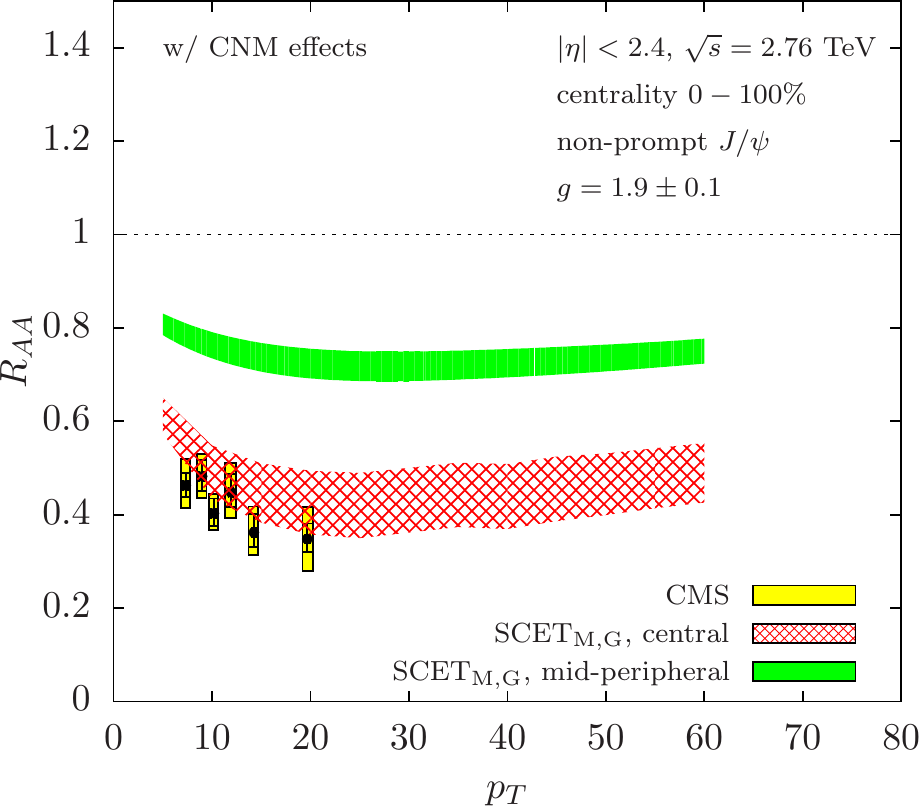} 
\end{center}
\vspace*{1.cm}
\caption{\label{fig:RAA-B-exp} Nuclear modification factor $R_{AA}$ for non-prompt $J/\psi$ production which originate from $B$-meson decays. The CMS data~\cite{Khachatryan:2016ypw} was taken at $\sqrt{s_\mathrm{NN}}=2.76$~TeV for $|\eta|<2.4$ and $0-100\%$ centrality. For comparison, we present our $B$-meson results for central ($0-10\%$ centrality, hatched red band) and mid-peripheral ($30-50\%$ centrality, green band) collisions. Again, we present our theoretical results without CNM effects (left, $g=2.0\pm 0.1$) and with CNM effects (right, $g=1.9\pm 0.1$).}
\end{figure*}

Finally, in Fig.~\ref{fig:RAA-B-exp} we compare to the $R_{AA}$ data from CMS~\cite{Khachatryan:2016ypw} for non-promt $J/\psi$ production which originate from the decay of $B$-mesons. The data was taken at $\sqrt{s_\mathrm{NN}}=2.76$~TeV for $|\eta|<2.4$ and 
is available only for minimum bias collisions ($0-100\%$ centrality). For comparison, we show our results for 0-10\% centrality and for mid-peripheral collisions with 30-50\% centrality. Again, we present our theoretical $B$-meson results without CNM effects (left, $g=2.0\pm 0.1$) and with CNM effects (right, $g=1.9\pm 0.1$). We find that our central results ($0-10\%$ centrality) agrees very well with the data. The minimum bias results are dominated by central collisions as they are weighted with the number of collisions. It will be instructive if the experiments can provide these data sets for fixed centrality bins, and thus to further test our theoretical framework. We look forward to more experimental data at the LHC in the near future.

\section{Conclusions and outlook \label{sec:four}}

We have derived a version of Soft Collinear Effective Theory that includes both the interactions with the medium that are mediated by Glauber gluon exchange and heavy quark masses. Using the new effective field theory, we obtained  vacuum and in-medium massive splitting functions for the  $Q\to Qg$, $Q\to gQ$ and $g\to Q\bar Q$ processes. Despite some ambiguities, we found agreement in the soft emission limit with earlier results in the literature where traditional approaches to parton energy loss in the QCD medium were used. In addition, we proposed a new formalism to include in-medium effects consistently at next-to-leading order in QCD. We  presented numerical open heavy flavor results for proton-proton collisions in the ZM-VFNS. Comparing with currently available data, we found good agreement even for relatively low $p_T$. Our numerical results for the suppression of open heavy flavor production in Pb+Pb collisions are applicable for $p_T\gtrsim 10$~GeV. We observed good agreement between theory and currently existing data sets for both $D$- and $B$-meson production at $\sqrt{s_\mathrm{NN}}=5.02$~TeV and 2.76~TeV. As it turns out, the low-$p_T$ suppression rates for open heavy flavor production are very sensitive to the relative contributions of heavy quark and gluon fragmentation. The currently available sets of fragmentation functions may not be sufficiently well constrained to make quantitative predictions in the very low $p_T$ region. In the future, there are several possible ways to improve the current framework. Firstly, our study clearly motivates global fits of heavy meson fragmentation functions which are currently only constrained from $e^+e^-$ data alone. Including both $pp\to HX$ and hadron-in-jet $pp\to (\mathrm{jet} H)X$ data will greatly improve the sensitivity in particular to the gluon-to-heavy-meson fragmentation functions. Secondly, it would be interesting to calculate the full in-medium splitting functions to second order in opacity. This way, it may be possible to extend the current framework down to lower $p_T$, where correlated multiple interactions with the medium become more relevant. This way, the full range of applicability of the current framework can be assessed and additional effects in the medium can also be taken into account.

\acknowledgments
We would like to thank Grigory Ovanesyan for his collaboration at the early stages of this work. In addition, we would like to thank Daniele Anderle, Yang-Ting Chien, Yen-Jie Lee, Emanuele Mereghetti, Rishi Sharma, Marco Stratmann, Hongxi Xing and Zhiqing Zhang for helpful discussions. This work is supported by the U.S. Department of Energy under Contract No. DE-AC52-06NA25396, in part by the LDRD program
at Los Alamos National Laboratory.

\bibliographystyle{JHEP}
\bibliography{bibliography}

\end{document}